\documentclass[lettersize,journal]{IEEEtran}
\usepackage{amsmath,amsfonts}
\usepackage{algorithmic}
\usepackage{algorithm}
\usepackage{array}
\usepackage[caption=false,font=normalsize,labelfont=sf,textfont=sf]{subfig}
\usepackage{textcomp}
\usepackage{stfloats}
\usepackage{url}
\usepackage{verbatim}
\usepackage{graphicx}
\usepackage{cite}
\begin{document}

\title{A Modeling of TSRCG and Resource Optimization for Multi-task Delivery Guarantee Algorithm Based on CGR Strategy in LEO Satellite Network}

\author{Xue Sun, Changhao Li, Lei Yan, and Suzhi Cao,~\IEEEmembership{Member,~IEEE}
\thanks{Xue Sun and Changhao Li are with University of Chinese Academy of Sciences, Beijing 100049, China, Key Laboratory of Space Utilization, Technology and Engineering Center for Space Utilization, Chinese Academy of Sciences,and also with Technology and Engineering Center for Space Utilization, Chinese Academy of Sciences, Beijing 100094, China (e-mail: sunxue16@mails.ucas.edu.cn, lichanghao20@mails.ucas.ac.cn).}

\thanks{Yan Lei and Suzhi Cao are with Key Laboratory of Space Utilization, Technology and Engineering Center for Space Utilization, Chinese Academy of Sciences, and also with Technology and Engineering Center for Space Utilization, Chinese Academy of Sciences, Beijing 100094, China (e-mail: yanlei@csu.ac.cn, caosuzhi@csu.ac.cn).}
}

\markboth{Journal of \LaTeX\ Class Files,~Vol.~14, No.~8, August~2021}%
{Shell \MakeLowercase{\textit{et al.}}:}


\maketitle

\begin{abstract}
With the reduction of satellite costs and the enhancement of processing capabilities, low earth orbit (LEO) satellite constellations can independently build inter-satellite networks without relying on traditional ground stations restricted by geographical distribution and can establish inter-satellite links   (ISLs) and complete computing and routing on-board. The characteristics of frequent on-off ISLs, the highly dynamic network topology of satellite networks make it face the challenges of routing strategy design as a delay/interruption tolerant network (DTN). As a deterministic dynamic routing algorithm, contact graph routing (CGR) uses a contact plan to calculate the path and forward data, but it still has problems such as high computational overhead, low prediction accuracy caused by ignoring queue delay, and overbooked problem caused by limited cache. Therefore, we first start with the time-space resource contact graph (TSRCG) to accurately characterize the time-varying and predictable characteristics of the satellite network and the network resource parameters under multi-tasks. Then, we optimize the route-list computation and dynamic route computation process to ensure task delivery and reduce the consumption of various resources, such as contact capacity, computing resources, and storage resources. And the resource optimization for the multi-task delivery guarantee algorithm based on CGR (RMDG-CGR) strategy we propose is compared with standard CGR in ION 4.0.1. Finally, the simulation results show that the RMDG-CGR can achieve higher task delivery in advance and successful task delivery rate, save contact volume occupancy rate, computing and storage resource, and the above effects are more prominent, especially in the task scenario with critical bundles.
\end{abstract}


\section{Introduction}
%
%
%
%
\IEEEPARstart{T}{he}  integrated air-space-ground network is one of the core directions of 6G and is listed as one of the seven critical requirements by the ITU. The space-based network composed of various satellites and satellite links is one of its essential subnets. The current satellite development already has on-board processing, switching, and routing capabilities, and does not have to rely on traditional ground base stations restricted by geographical distribution. Multiple satellites can have inter-satellite links(ISLs) and form constellations. At present, various countries are accelerating the development of satellite Internet, such as SpaceX’s Starlink Project, OneWeb and Amazon’s Kuiper, Canada’s Telesat low-orbit broadband constellation, the Russian Aerospace State Group (ROSCOSMOS)’s "Sphere" constellation program, SpaceNet by India's Astrome Technologies and China Aerospace Science and Technology Corporation's low-orbit communication satellite constellation "Hongyan". In recent years, the significant reduction of satellite manufacturing and launch cost and advanced mobile communication technology have provided a technical guarantee for the rapid development of satellite Internet. Tasks such as environmental monitoring, intelligence reconnaissance, military applications for emergency rescue and disaster relief, and future access to new services for personal terminals pose challenges to satellite networks. However, the delay and interruption conditions in space hinder the effective implementation of traditional Internet protocols, which rely on the frequent handshake feedback information. Therefore, the US Jet Propulsion Laboratory (JPL) proposed a network architecture called delay/disruption tolerant network(DTN) for the development of Inter Planetary Internet(IPN) to overcome these problems, and it has been widely used in ground networks such as military combat networks and sparse sensor networks. DTN is an optional solution for constructing satellite networks\cite{r1}, especially to meet the intermittent connection needs of low-earth orbit(LEO) constellation systems\cite{r2}, that is, signal propagation delay is not the main limiting factor in LEO scenarios. Disruptions due to orbital dynamics, highly directional antenna orientation or platform power limitations are the problem. Although the LEO satellite constellations have periodic changes, there are still some insurmountable challenges, such as limited transmission bandwidth and buffering, continuous dynamic changes in network topology, variable data flow distribution, and high bit error rate. With the increase of the constellation scale, the number of ISLs increases sharply, which leads to an exponential increase in the related costs and risks. Based on the above challenges, it can be seen that for the inter-satellite dynamic network environment, it is necessary to develop a dedicated routing algorithm that should calculate the route with lower communication volume and computation overhead and adapt to the dynamic satellite network topology in real-time.

DTN routing technology can be divided into two categories according to the amount of knowledge used in decision-making, one is opportunistic routing that does not use any network information, and the other is deterministic routing that uses perfect knowledge of the network. The former is usually based on the flooding strategy. Although it reduces the complexity of the algorithm, too many message copies are a huge consumption of network resources and are not suitable for satellite networks with limited resources\cite{r3}\cite{r4}. The latter is suitable for DTN networks where the connection time between nodes is known or can be accurately predicted, such as the LEO satellite scenario that this article focuses on. NASA JPL designed a dynamic routing algorithm called contact graph routing(CGR)\cite{r5}, it is a dynamic routing system that computes routes through a time-varying topology of scheduled communication contacts in a DTN network. Deterministic routing has been widely studied. Time-independent graph routing(TIGR) models DTN as an equivalent time-independent graph and uses general routing algorithms to obtain the best results based on time-independent graphs\cite{r6}; least delay routing (LDR)\cite{r7}represents the network as an undirected contact graph, using the modified shortest path algorithm; the earliest arrival optimal delivery ratio(EAODR)\cite{r8}routing selects the earliest arrival path from a given node, the node checks all paths with the required earliest departure time and the future; the routing strategy based on time expansion graph(TEG) aimed at improving QoS \cite{r9}\cite{r10}abstracts the dynamic network into a graph-based maximum commodity flow problem. Most of the above algorithms assume that the data packet will be sent at the beginning of the contact. There is no careful and precise consideration of the queuing delay of the data in the routing process. In DTN, even a tiny delay error may cause unplanned connection failure\cite{r11}.

Among the routing algorithms suitable for DTN mentioned above, CGR has received more and more attention in recent years and has been able to prove sufficient accuracy and efficiency to become the de facto routing framework for spatial DTN\cite{r12}. The CGR core algorithm uses the Dijkstra algorithm based on the contact graph to select the best path and takes the estimated delivery time (EDT) as the optimization goal. However, NASA officially implemented the CGR algorithm in its open-source software ION and did not consider the queuing status of nodes due to the storage function of DTN at first, so \cite{r13} proposes to use the earliest transmission opportunity (ETO) parameter to estimate the actual transmission time of the data.\cite{r12} describes how to implement the construction of the candidate route list in the route search algorithm. Two additional procedures are added on the basis of EDT and ETO, namely the projected arrival time (PAT) and the effective volume limit (EVL).

In addition, since DTN strongly relies on a contact plan (CP) in actual practice, it is also an essential input for implementing the CGR algorithm. So it is more effective for a small-scale network with several nodes, and it will be difficult to extend to large-scale networks in the future, such as LEO constellations\cite{r14}, and the overlapping of contacts will be more complicated than in the deep space environment. Therefore, it is necessary to build a more accurate and strict unified mathematical model or framework to describe the DTN network applied in space so as to deal with the subsequent analysis of complex and large-scale networks, services, architecture protocols, and software.

Summarizing the above-related work, the current CGR research under LEO constellations has the following problems: (1) Each intermediate node on the CGR path will recalculate the optimal forwarding path of the bundle to cope with changes in the network state. Recalculation brings flexibility to the algorithm while also increasing the computational cost of the algorithm\cite{r15}. (2) The EDT of a bundle should be affected by various network factors such as its own priority, survival time, node cache amount, contact volume, etc. In complex scenarios where the end-to-end links are frequently interrupted and overlap each other , standard CGR will waste some contact transmission opportunities with higher bandwidth/rate. How to describe the EDT more accurately to ensure the delivery rate of various tasks\cite{r16}. (3) It is necessary to build a more precise and rigorous unified mathematical model or framework to describe the DTN network applied to the space so as to deal with the subsequent analysis of complex and large-scale networks and services, architecture protocols and software. (4) In order to evaluate the effectiveness of DTN technology as a mitigation of these challenges, we have to face up to the problem of requiring a large amount of local high-speed memory to accommodate a large number of bundles\cite{r18}.

In order to overcome the above bottleneck of CGR, we need to build a more comprehensive satellite network dynamic model and also consider the generation of the routing list and the forwarding mechanism of the bundle to ensure the delivery rate and resource consumption of different types of tasks. The contributions of this article are summarized as follows:

\begin{enumerate}
\item{We propose the time-space resource contact graph (TSRCG) model to describe the network link status and various resource conditions (the contact volume, node computing resources, node storage resources, and other knowledge) in the two dimensions of time and space, and accompanied by a multi-task model input that can have up to 3 types of bundle priorities, including the start/target node of the task, bundle size, priority, arrival time and expiration time, etc.}
\item{Using the task model based on TSRCG, combined with the satellite tool kit (STK) and Python tools, it is easy to adapt large-scale LEO constellations of any configuration as a scenario for experimentation. In addition, the task transmission status, link status, and node status information can be grasped during the task transmission process.}
\item{After obtaining the above TSRCG and task model, we propose the resource optimization for multi-task delivery guarantee algorithm based on CGR (RMDG-CGR) algorithm. Specifically, it modified the route-list computation and dynamic route computation in the bundle routing process, that is, the cost function based on EDT in the forwarding process and the optimization of the routing allocation mechanism for critical bundles.}
\item{The theoretical and simulation analysis of the models and algorithms proposed above have verified the effectiveness of the proposed schemes in this paper.}

\end{enumerate}

The rest of this article is organized as follows. The Section \uppercase\expandafter{\romannumeral2} introduces TSRCG and the multi-task model. In the Section \uppercase\expandafter{\romannumeral3}, we propose and describe in detail the RMDG-CGR algorithm based on TSRCG. Then, in the Section \uppercase\expandafter{\romannumeral4}, we discussed the simulation results. Finally, conclusions are drawn in Section \uppercase\expandafter{\romannumeral5}.

\section{System model}
In this section, we first describe the TSRCG model that is applicable to LEO satellite constellations network that implements the CGR method based on the DTN network and considers various constraints. Then, we combine the model with multi-tasks.

\subsection{The TSRCG Model}
CGR is a distributed path calculation method. It assumes that the network has topology pre-calculation capabilities, and each node in the network can obtain global network contact information and local queue occupancy in a timely and accurate manner. It relies on the CP to generate a network contact graph. Contact refers to the opportunity to establish a communication link between two DTN nodes. CP refers to a timing list composed of all feasible contacts within a topological interval\cite{r19}. The CGR strategy can be realized based on the following important premise: the communication plan of the nodes in the network is inferred, and each node knows the overall situation, that is, the plan of other nodes, and the routing information is not discovered through the traditional TCP/IP three-way handshake dialogue but is obtained directly through the CP. The CP in the standard CGR consists of two types of messages, one is a contact message, and the other is a range message. Among them, the contact message includes the start time, end time, sending node, receiving node, and data transmission rate (B/s) of the contact. The range message contains the start time, end time, sending node, receiving node, and the distance(in light-seconds)  between the sending node and the receiving node of the contact.

Based on the contact graph, we give the time-space and resource characteristics of the LEO satellite constellation network, such as the contact volume resources, node computing resources, node storage resources, and other factors under the predictable time-varying topology, and named this graph TSRCG. First, we introduce the main parameters in the TSRCG satellite network. We define $ C_{i,j}^{t_1,t_2} $ to be expected during a time interval $(t_1,t_2)$ that satellite $i$ can send data to satellite $j$. In addition, we also define the rate $R$ and the distance $D$ to represent the data transmission rate, and the distance between the satellites, the calculation method of the distance is detailed in\cite{r20}. After we use STK and MATLAB to obtain the actual inter-satellite distance, according to the standard CGR protocol, we also calculated the one-way light time (OWLT) in the TSRCG, which is the distance of light seconds. Due to the high mobility between the two satellite nodes of the space network, for the LEO Walker configuration constellation that this article focuses on, especially the situation between two different orbit satellites. Although it meets the limit of the range of communication distance, the relative distance is still constantly changing. The meaning of OWLT margin is the maximum increment of OWLT between any pair of satellite nodes that can be received by defining a safety margin for the node distance during the transmission of the bundle\cite{r21}. According to the parameter $40$ miles/sec of Helios, the fastest artificial spacecraft so far, as an example, if any two nodes move relative to each other in completely opposite directions, the distance increases by up to $80$ miles/sec. Therefore, the total transit time of data can be estimated in the most pessimistic case as
\begin{equation}
T=D+2Q
\end{equation}
Among them, $D$ is the number of light-seconds between the two satellites, $Q$ is the OWLT margin and
\begin{equation}
Q=40\times\frac{D}{18600}
\end{equation}

Then, in addition to showing the topological characteristics that change over time, we also express the following expressions for contact volume resources, node routing computing resources, and node storage resources in TSRCG. The calculation method not only follows the specific implementation process of the standard CGR algorithm in ION, but also reflects the actual utilization of the entire satellite constellation network resource under the long-term constellation period.

\subsubsection{Contact Volume Resource ($C.Volume$)}
At the current time $t$, we filter the transmission start time $t^C_S$ and the end time $t^C_E$ of all contacts $c\in C$ to include the contacts at time $t$, and we can get the set of available contacts $\hat{C}_t$:
\begin{equation}
\hat{C}_t=\{c|c\in C,t^C_S\leq t^C_E\}
\end{equation}
Among them, in the set $\hat{C}_t$, the contact volume resource $V_C$, namely $C.Volume$, is obtained by checking and updating the contact that is carrying out the bundle transmission task at the current time $t$.
\begin{equation}
V_C=|\{c|c\in \hat{C}_t,\mbox{$c$ contains bundles transferring in progress}\}|
\end{equation}

From the above, we can also show that the contact volume occupancy rate at time $t$ is
\begin{equation}
R_O=\frac{V_C}{|\hat{C}_t|}
\end{equation}

\subsubsection{Satellite Node Routing Computing Resource ($S.Computing$)}
This variable is defined as the number of execution iterations of the two functions Yen-PLUS and Candidate Route Construction in the RMDG-CGR algorithm we proposed in Section \uppercase\expandafter{\romannumeral3}.

\subsubsection{Satellite Node Storage Resource ($S.Storage$)}
We define the number of bundles cached at time $t$ on a certain satellite node $V$ as $S.Storage$. The source node of these bundles may be the current node, or these bundles may also be in the forwarding queue of the current node waiting for the next transmission opportunity. It is also possible that the target node of these bundles is the current node that has reached the destination. This indicator can help measure and judge the setting and consumption of on-board storage resources in future engineering practices.

\begin{figure*}
    \centering
    \subfloat[]{\includegraphics[width=0.55\textwidth]{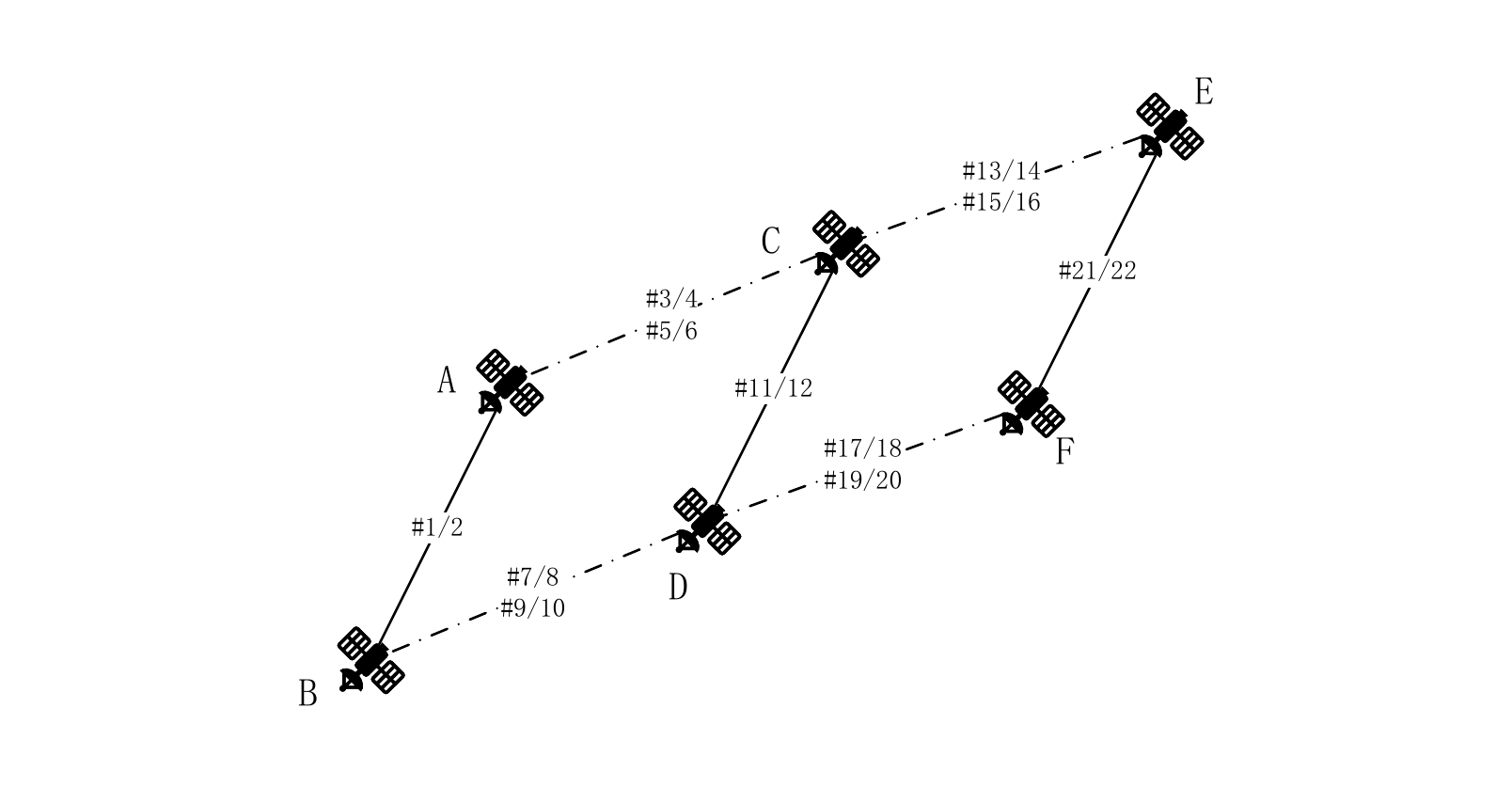}%
    }
    \hfil
    \subfloat[]{\includegraphics[width=0.4\textwidth]{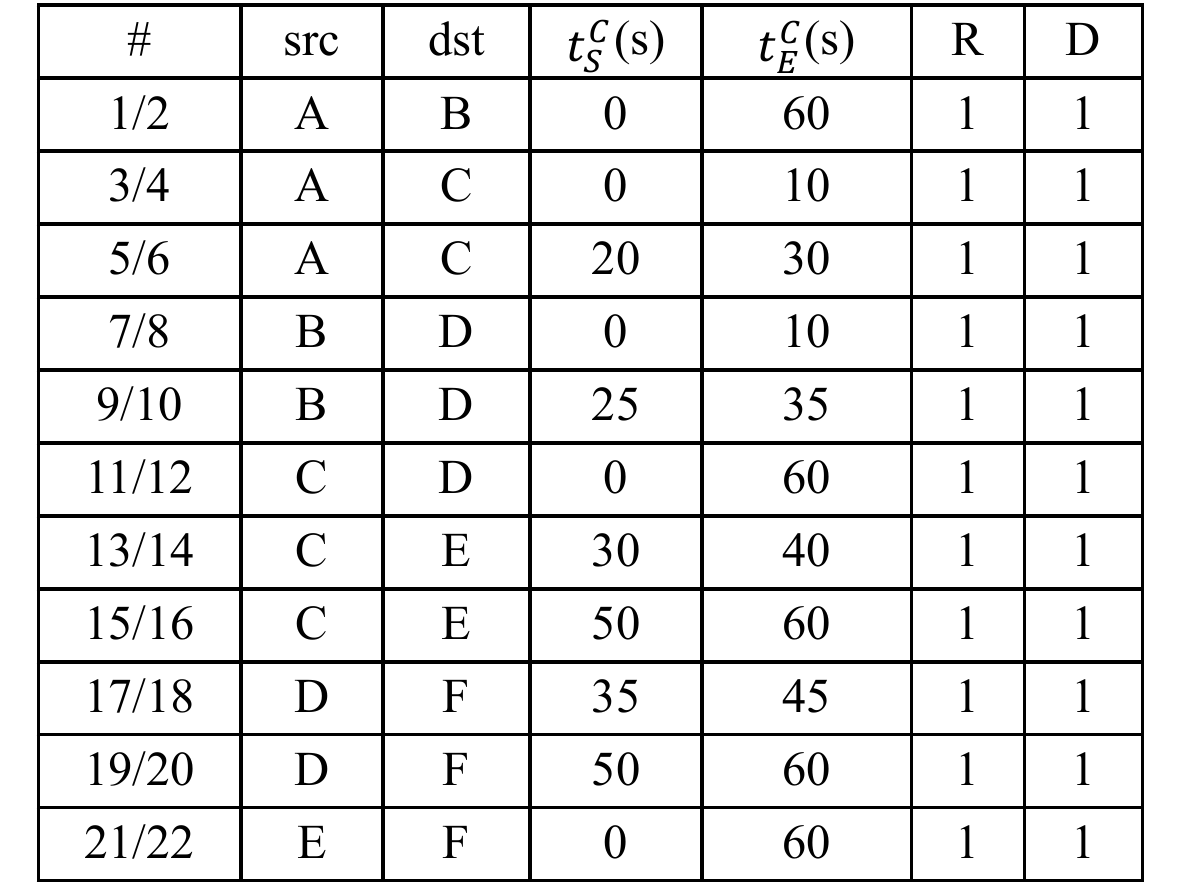}%
    }
    \caption{An example of Delta LEO satellite network represented by (a)static graph of the topology, (b)contact plan table.}
    \label{Fig1}
\end{figure*}

In Fig. \ref{Fig1} (a), we take the classic Delta configuration LEO constellation as an example (the constellation period is T) to show the traditional static graph's representation of the physical location of the node. Still, it cannot provide an intuitive understanding of the time dynamics. The table in Fig. \ref{Fig1} (b) lists each contact $C$ identified by a number (\#1$\sim$16), among which, because there are permanent links between adjacent satellites in the same orbit, $ C_{A,B}^{0,T} $, $ C_{C,D}^{0,T}$ and $C_{E,F}^{0,T} $ stand for permanent links. The other links represent episodic ISLs according to the continuous distance change between the satellites. According to the definition of CGR, the contact is one-way, so we use a pair of one-way contacts to represent two-way communication. Unlike deep-space and satellite-to-earth environmental communications, we consider that the two-way communication rate and distance between LEO satellites are symmetrical. For ease of presentation, in this example, we assume that R is 1 and D is 1, and the constellation period T is assumed to be 60s.

\begin{figure*}[!t]
\centering
\includegraphics[width=0.7\linewidth]{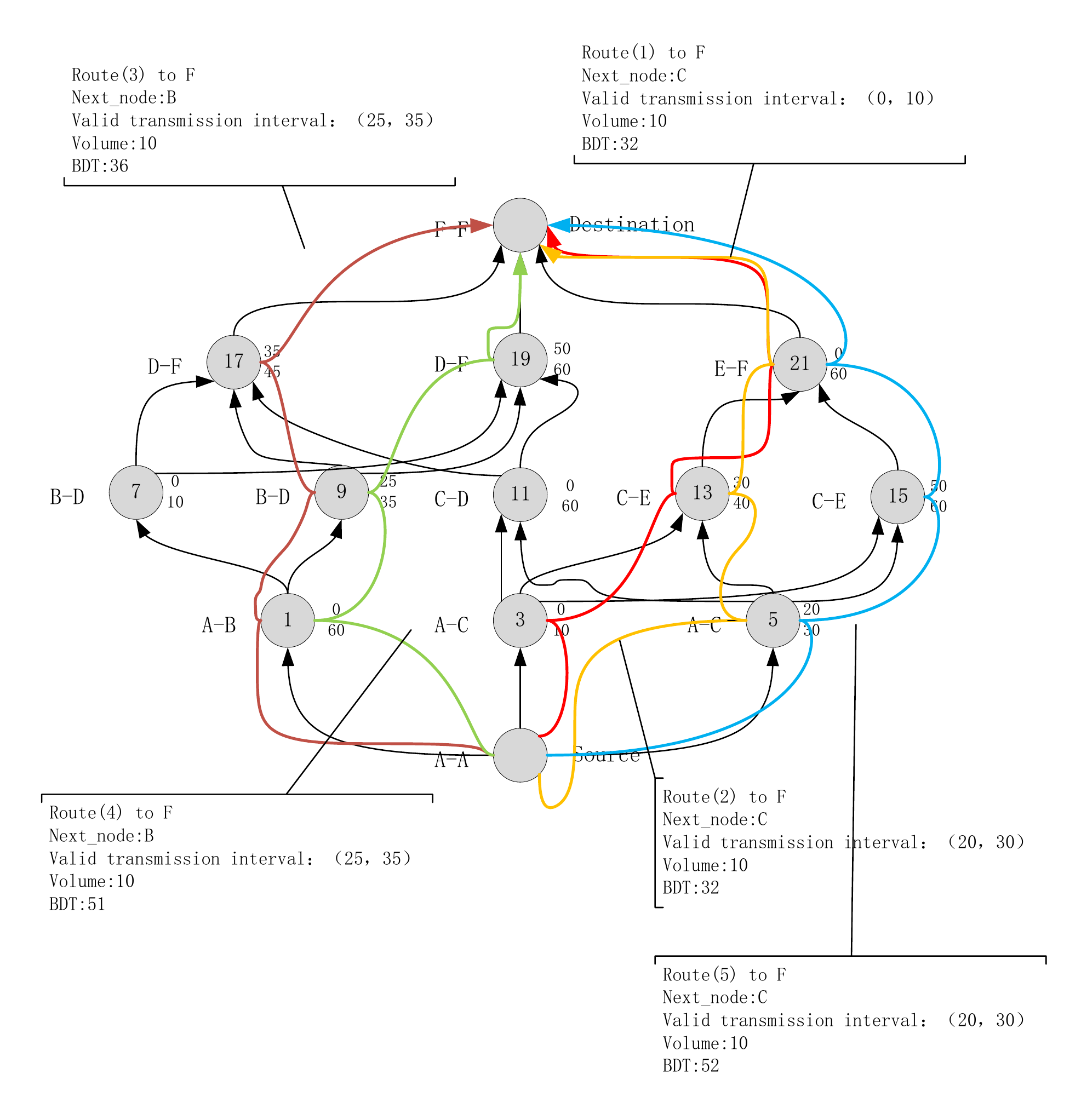}
\caption{Time space resource contact graph $TSRCG_A^F$.Example routes from A to F are highlighted.}
\label{Fig2}
\end{figure*}

According to the above definition, we get a directed acyclic graph $TSRCG_S^D=\{(V,E,T,C_{u,v}^T,V_C,S.Computing,\\S.Storage)\}$where $u$ and $v$ are the satellite node labels, and $T$ is the time range of the contact. The structure of this graph seems to have almost nothing to do with the physical network topology shown in Fig. \ref{Fig1}, but it overcomes the shortcomings of other models, such as traditional static graph representation\cite{r22} and time-evolving graphs\cite{r23}\cite{r24}. The TSRCG can reflect the time dynamics, scalability, and accuracy of the satellite network topology throughout the cycle and can characterize space resources such as contact volume limitations and node routing computing resource consumption, which will help us to execute the network algorithm in Section \uppercase\expandafter{\romannumeral3} on this basis. Fig. \ref{Fig2} illustrates the TSRCG from the source node A to the destination node F based on the network topology shown in Fig. \ref{Fig1}, where the vertex $V$ corresponds to the contact, which is the contact time period during which data can be transmitted, and the edge $E$ is not the path to achieve data transmission, but the time period during which the bundle must be stored in this node while waiting for subsequent contact.

In Fig. \ref{Fig2}, we use colored lines with different colors to show several candidate routes(not all) from the source node A to the destination node F. The valid transmission interval (VTI) is determined by the earliest and latest time when node A can start transmitting data to destination node F.Volume means the product of contact duration and transmission rate. The best delivery time (BDT) is the earliest possible time that the first byte of the data can reach the target node after considering the contact propagation delay and the storage time caused by the temporary unavailability of the next-hop link, that is, the best time to start delivering data. For example, if a bundle wants to transmit from A to F in VTI of (0,10), $Route(1)R_A^F=\{C_{A,C}^{0,10},C_{C,E}^{30,40},C_{E,F}^{0,60}\}$ is the fastest route, and choosing this route means that the maximum amount of data that can be carried from $A$ to $F$ is 10. During this period, the BDT value is the smallest.

\subsection{The Multi-task Model}
Earlier, we have modeled the TSRCG in the LEO satellite constellation network scenario. In the case of having mastered the contacts and routes, after completing the routing computation and storing it in the routing table, the data is about to be forwarded through various intermediate nodes, which means that the available resources on the contacts and nodes are being consumed. If the available contact volume or storage capacity is insufficient, the bundle may be blocked from forwarding, thus affecting routing decisions and final delivery. According to the classification of service of bundle data, it has different priority policies in forwarding.

In the CGR implementation based on the 4.0.1 version of ION, the forwarding and delivery strategy for bundles with different priorities is that if the bundle is marked as critical, which is the highest priority, a copy of it will be transmitted to each candidate route list neighboring nodes to ensure as far as possible that it can be successfully transmitted to the final destination. For non-critical bundles, the CGR dynamic routing algorithm will select a single adjacent node to forward the bundle, but due to some unforeseen delays, the selected node may prove to be a sub-optimal forwarder, then the arrival time of this bundle may not be optimal or even unreachable.

In the CGR-based route information, the key information we pay attention to includes expiration time and whether it is a critical bundle. Among them, the expiration time is the attribute of each bundle, and its impact on routing is reflected in the fact that when the local node calculates the next-hop node for a bundle, it needs to check whether the bundle can arrive before expiration. If not, it cannot choose the path.

\begin{equation}
\mbox{Expiration time=Generation time+TTL(time-to-live)}
\end{equation}

To more truly reflect and verify the priority strategy of tasks in the routing and forwarding process, we have conducted research on three representative types of data\cite{r16}, which are streaming data such as satellite telemetry data, expedited data such as emergency sensor data, and bulk data such as images. And three different types of bundle tasks are characterized by eight attributes: $M_i=\{B_{ID}(M_i),S(M_i),D(M_i),P(M_i),Flag_{critical}(M_i),$ $T_G(M_i),T_{Ex}(M_i)\},i=1,2,...,n$, where $B_{ID}(M_i)$ represents the label of the bundle, $S(M_i)$ and $D(M_i)$ are the source and destination satellite node of $M_i$ respectively,$B(M_i)$ represents the size of bundle, $P(M_i)$ is the priority of $M_i$, $Flag_{critical}(M_i)$ characterizes whether the $M_i$ is critical, and $T_G(M_i)$ and $T_{Ex}(M_i)$ are generation time and expiration time of $M_i$ respectively.

The routing strategy proposed in this paper strictly complies with the CCSDS's provisions on the priority issue in bundle routing\cite{r25} when queuing forwarding on each node. It is assumed that only after each bundle with a higher priority that is queued for transmission to the same neighbor has been transmitted the bundle can be transmitted to the neighbor. Each bundle has a priority, and different priorities correspond to a service level. The priority determination method in this paper is specifically defined as the following three types. Among them, the lowest priority is assigned to the bundle with service level 0, and the highest priority is assigned to the bundle with service level 2. This determination method can help optimize contact utilization.

\subsubsection{Streaming Traffic}
We first consider the streaming data regularly generated in the satellite network, which can usually be understood as telemetry data. It is assumed that a new 1Mb bundle is generated every 5s, the priority is 2, and $Flag_{critical}(M_i)$ is set to 1. The transmission requirements are as fast as possible and must be reliable because the routing strategy we will propose below will guarantee the delivery of such tasks as quickly as possible with a delivery rate of $100\%$.
\subsubsection{Expedited Traffic}
The size of a single bundle of this type of task is between 1 and 5Mb, and a maximum of 3 are generated every 10s. This can usually be understood as emergency sensor data. The priority set in the model in this paper is 1, and $Flag_{critical}(M_i)$ is set to 0.
\subsubsection{Data Traffic}
Data is similar to an image. For example, a burst of 20 bundles (each size between 1 and 5Mb) is generated within 25s, and their priority is the lowest 0. In this model, $Flag_{critical}(M_i)$ is set to 0.

\section{The RMDG-CGR Strategy}

\subsection{Overall Routing Scheme}

\begin{figure}[!t]
\centering
\includegraphics[width=0.9\linewidth]{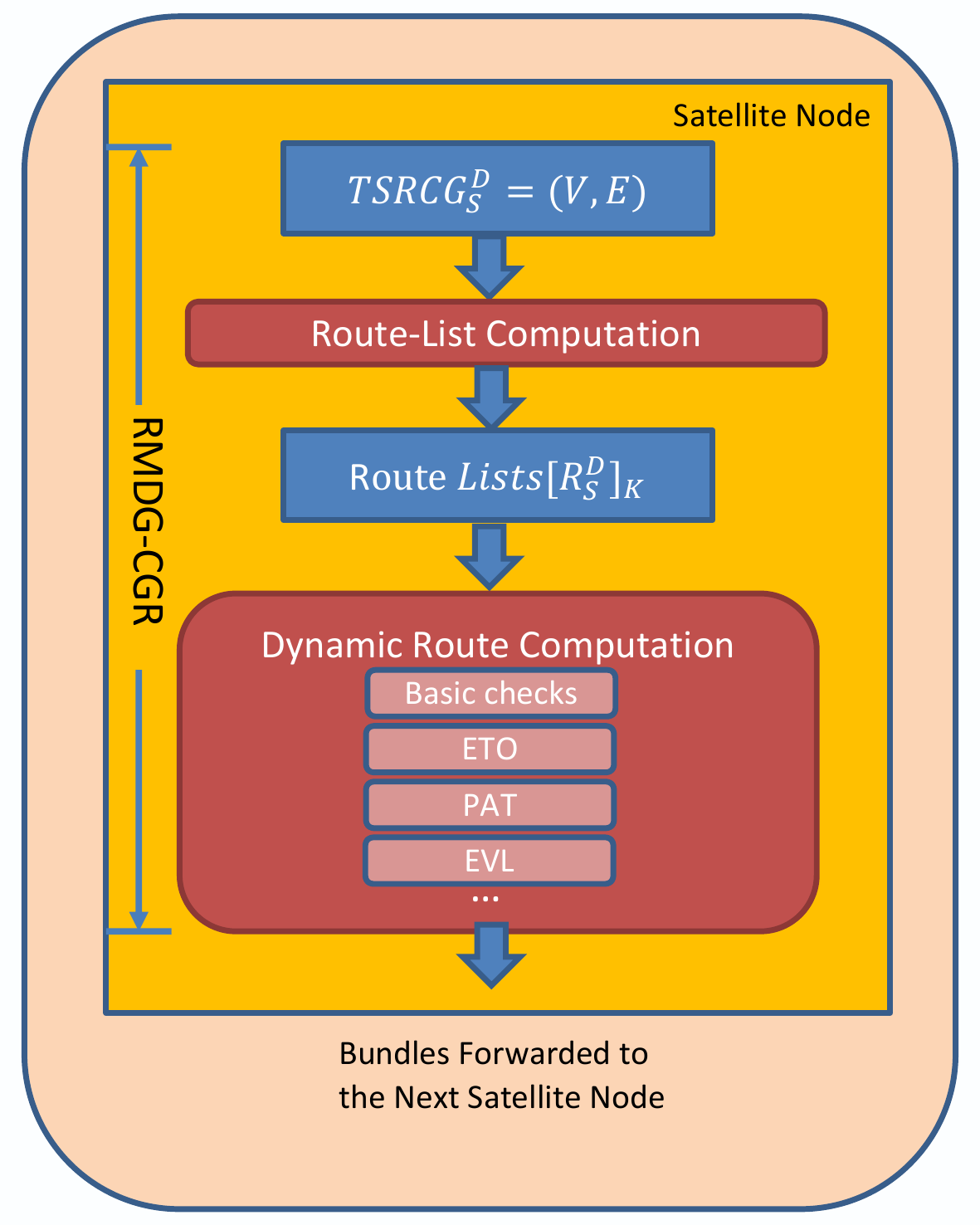}
\caption{The overall operation and processing of RMDG-CGR strategy for bundle delivery.}
\label{Fig3}
\end{figure}

First of all, the successful operation of RMDG-CGR relies on each DTN node participating in the forwarding to have accurate TSRCG as input. As shown in Fig. \ref{Fig3}, TSRCG-based routing calculations are completed inside each satellite node, including route-list computation, route review, route selection, and queuing. This process will be repeated at each downstream satellite node until the bundle is successfully delivered to the destination. Route-list computation is the core of the construction of the subsequent candidate route list and the path selection of the bundle based on the input of the specific task. Dynamic Route Computation follows four check steps\cite{r12} to determine the final path for a bundle for forwarding, which are basic checks, ETO, PAT, EVL, etc. When the bundles forwarding program is executed between satellite nodes, the current node will refer to the current local queue status information and the bundle parameters to be forwarded to select the best candidate route for the destination and place it in the outbound queue for transmission.

In 3.6 and earlier versions of ION, distributed static routing is implemented, which means that all routes from the source to the destination node are calculated at one time\cite{r27}. The disadvantage is that it is required in a large-scale satellite constellation network with limited resources. The calculation takes a long time, and once the CP is changed, the previous calculation results will be cleared and recalculated (occupying forwarding time). The whole strategy process of RMDG-CGR proposed in this paper is a distributed dynamic routing algorithm, which is consistent with the latest version of the ION function. We use TSRCG to calculate a limited number of routes and then expand and update the route-list according to the task traffic demand of the local node and the bundle queuing status. So each satellite node under dynamic routing can combine its own calculation and storage resource conditions to calculate a limited number of new optimal routes on demand.

\begin{figure*}[!t]
\centering
\includegraphics[width=0.95\linewidth]{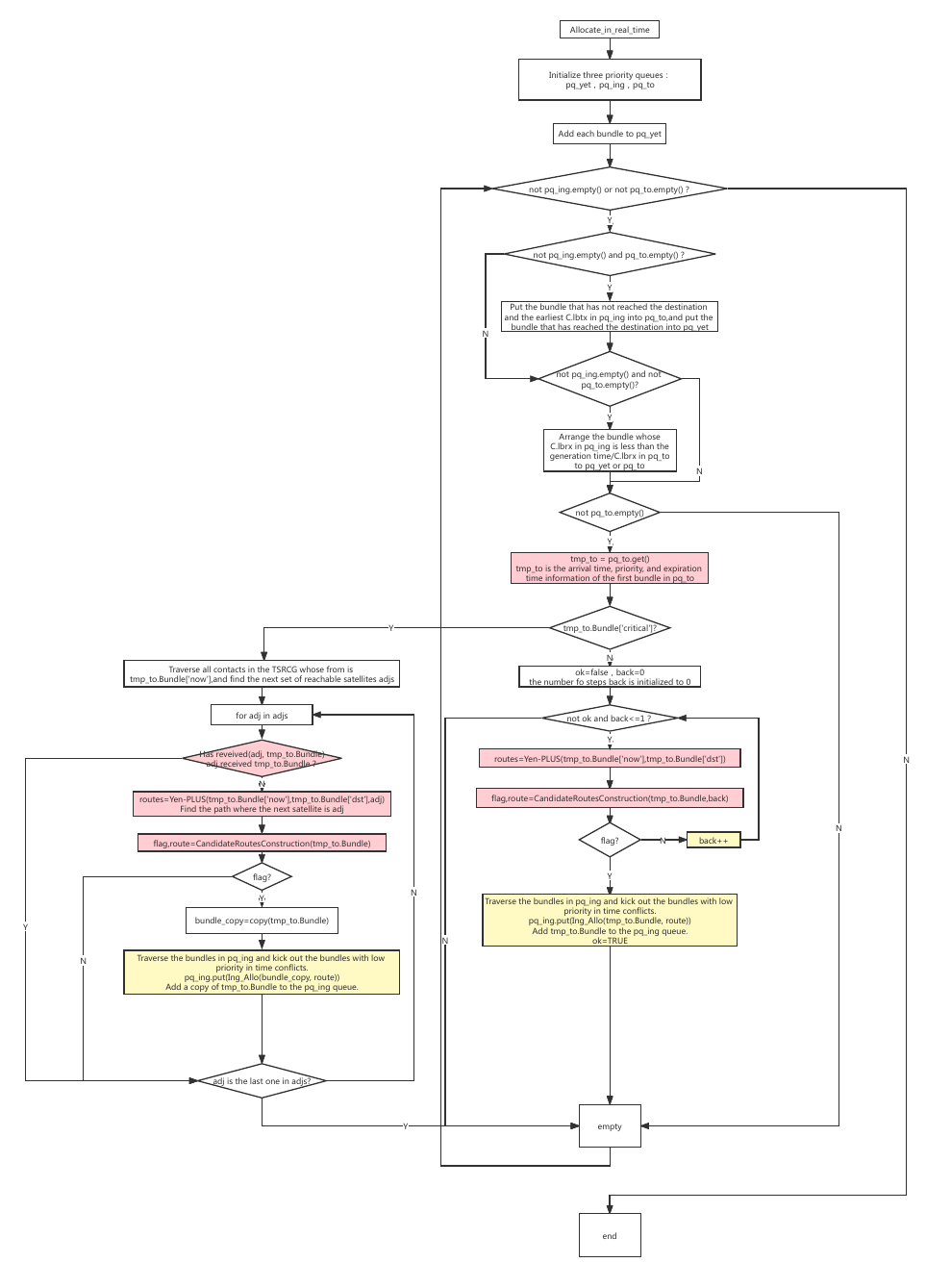}
\caption{RMDG-CGR algorithm flowchart.}
\label{Fig4}
\end{figure*}

The algorithm flow chart of RMDG-CGR is shown in Fig. \ref{Fig4}, in which the red and yellow highlighted parts represent the core operations in the algorithm implementation. We will follow the steps in B, C, and D below to describe the specific implementation, innovation, and complexity of the algorithm.

\subsection{Route-List Computation}
The purpose of this process is to calculate the K shortest path route $[R_S^D]_K$ based on TSRCG to ensure the delivery of bundles with different priorities, expiration times, and data source modes.

Given the TSRCG and the multi-task model, the path search algorithm we propose is a problem of finding the shortest path between all pairs of vertices in the graph. Considering the actual situation of the overbooked problem that the bundle priority may cause, we not only hope to get only one best path, but also hope to get multiple paths such as sub-optimal and sub-sub-optimal for decision-making reference. Therefore, it is necessary to expand and extend the shortest path problem based on Dijkstra into the issue of searching for K shortest paths (KSP). Since Yen adopts the deviating path algorithm idea in the recursive method, it is suitable for the KSP solution in the directed acyclic graph with non-negative weight edges. Therefore, our algorithm is further optimized and improved based on Yen's Algorithm\cite{r28} implemented for CGR in the ION 4.0.1 version, which is called Yen-PLUS.

Suppose that the result of the path search from a source node $S$ to a target node $D$ based on TSRCG is $[R_S^D]_K=\{[R_S^D]_1,[R_S^D]_2,[R_S^D]_3,...,[R_S^D]_k\}$. The K paths should have the following attributes:(i)K paths are generated in order, that is, for all $i(i=1,2,...,K-1)$, $[R_S^D]_i$ is determined before $[R_S^D]_{i+1}$.(ii)K paths are arranged in descending order of cost function value, that is, there is $cost([R_S^D]_i)<cost([R_S^D]_{i+1}$ for all $i(i=1,2,...,K-1)$.(iii)The k paths are the shortest, that is, there is $cost([R_S^D]_K)<cost([R_S^D])$ for all $[R_S^D]\in[R_S^D]_{SD}-[R_S^D]_K$.

First, according to the given contact plan CP, the source S and the destination node D use the Dijkstra algorithm to calculate the shortest path from S to D, record it as $P_k(k=1)$ and add it to the shortest path set $[R_S^D]$(this step can be seen for initialization) and then judge whether the number of paths k has been found is less than the number of K paths we need and whether there are still potential paths in the potential path list $[P_S^D]$. If not, the K shortest paths have been found, and the program ends; if yes, continue to search for the $P_{k+1}$ path, starting with each node on $P_k$ (excluding the target node $D$) as the spur node $V_i$ and traversing in turn, finding the shortest path from $V_i$ to $D$(that is, the second Dijkstra algorithm is executed). It should be noted that to prevent the overall path from the source node to the destination node from being looped, the shortest path from $V_i$ to $D$ cannot include any node on the path from the source $S$ to the spur node $V_i$. Moreover, to avoid duplication with paths already in the result list, the edge starting from $V_i$ cannot be the same as the edge starting from $V_i$ on the route included in the result list $[R_S^D]_K$. When traversing spur nodes, the paths found through the above process and constraints are called potential routes, and they are all added to the potential list set $[P_S^D]$. If $[P_S^D]$ is not empty, find the path with the least cost and move it to $[R_S^D]$, which is $P_{k+1}$.

Generally, BDT can be used as the path cost to find the shortest path from the source satellite node to the destination satellite node in the contact graph\cite{r12}. However, when $[P_S^D]$ for a certain destination satellite contains multiple routes, in this paper, we define the cost function as the EDT, and the idea of determining the best route for EDT is the route with the smallest BDT among all the potential routes. If there are multiple routes with the same and smallest BDT value, select the least number of contacts, and if there are still routes with the same conditions, select the latest end time of VTI. If the above conditions are the same, the route with the smallest node number will be selected last. The above path searching and sorting process is called Yen-Plus, and the detailed calculation method of EDT is introduced as follows.

When the original Yen algorithm is executed, we obtain the set of $K$ paths from $S$ to $D$ as $\mathbb{R}$. Next, we calculate the respective EDT for these K paths. The EDT of the i.th path is expressed as follows:
\begin{equation}
   \begin{split}
   \mathbb{R}_i.EDT=&\mathbb{R}_i.BDT\times\omega^3+\mathbb{R}_i.{hop\_cnt}\times\omega^2\\
   +&\mathbb{R}_i.VTI\times\omega^1+\mathbb{R}_i[1].index
   \end{split}
\end{equation}
Among them, $\mathbb{R}_i.BDT$ is the BDT from $S$ to $D$ through the path $\mathbb{R}_i$, $\mathbb{R}_i.{hop\_cnt}$ represents the number of hops from $S$ to $D$ through the path $\mathbb{R}_i$, $\mathbb{R}_i.VTI$ represents the valid transmission window of the contact from $S$ to $D$ through the path $\mathbb{R}_i$, and $\mathbb{R}_i[1].index$ represents the number of the first hop of $C$, $\omega$ is a factor used to quantify the weight. Here, $\omega$ takes 999999999 according to the situation.

\begin{figure*}
    \centering
        \subfloat[]{\includegraphics[width=0.31\textwidth]{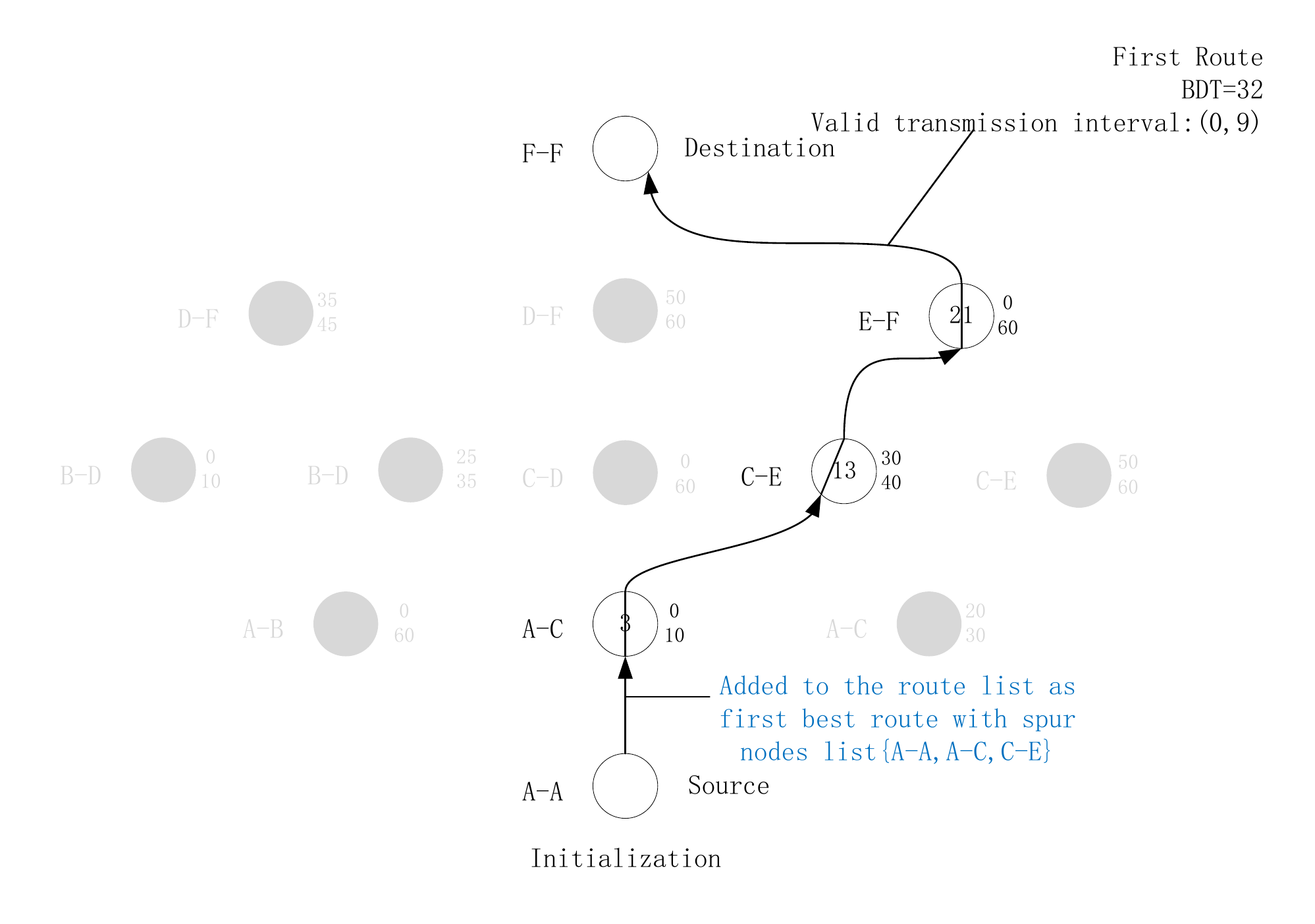}%
    }
    ~
        \subfloat[]{\includegraphics[width=0.31\textwidth]{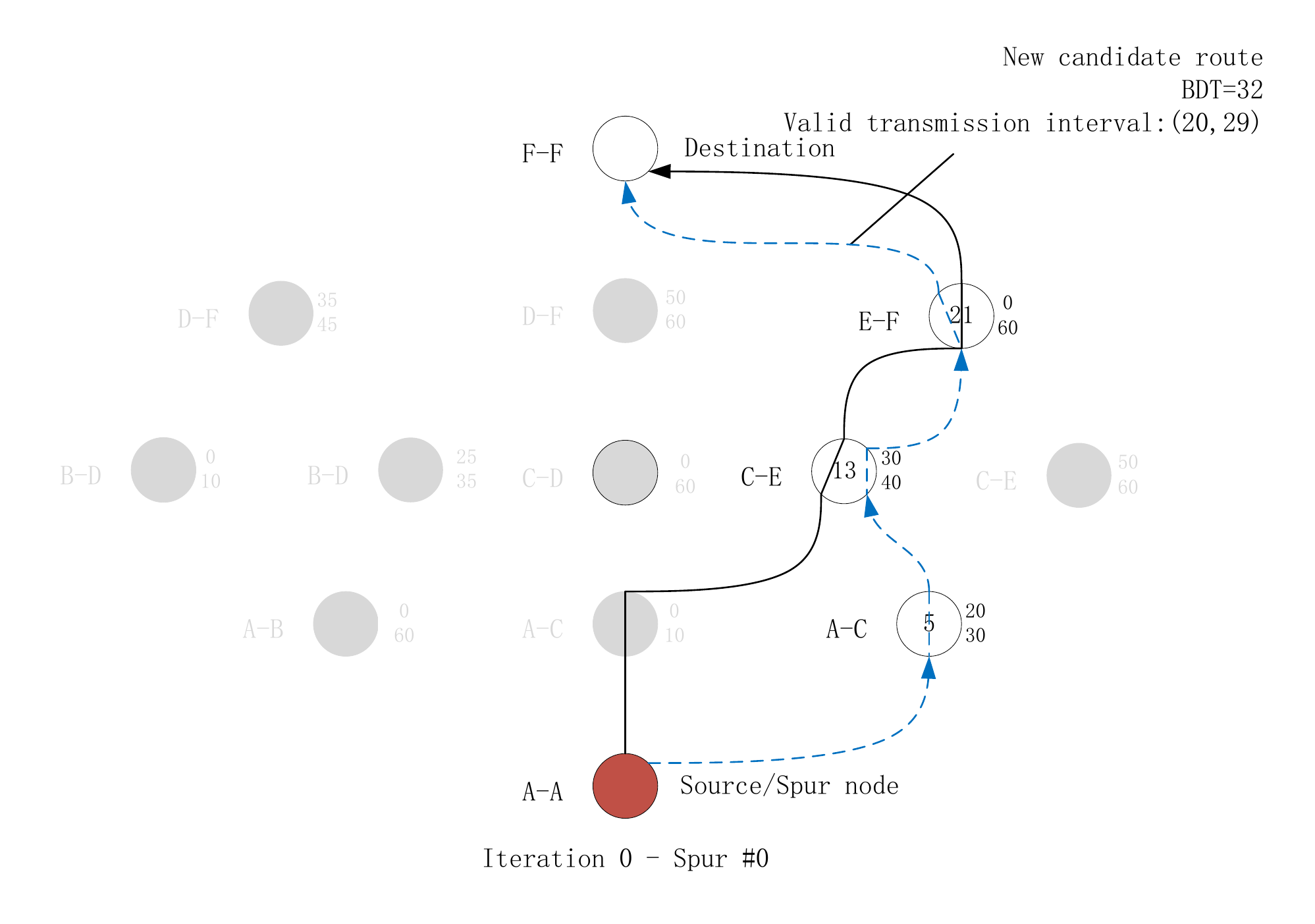}%
    }
        ~
        \subfloat[]{\includegraphics[width=0.31\textwidth]{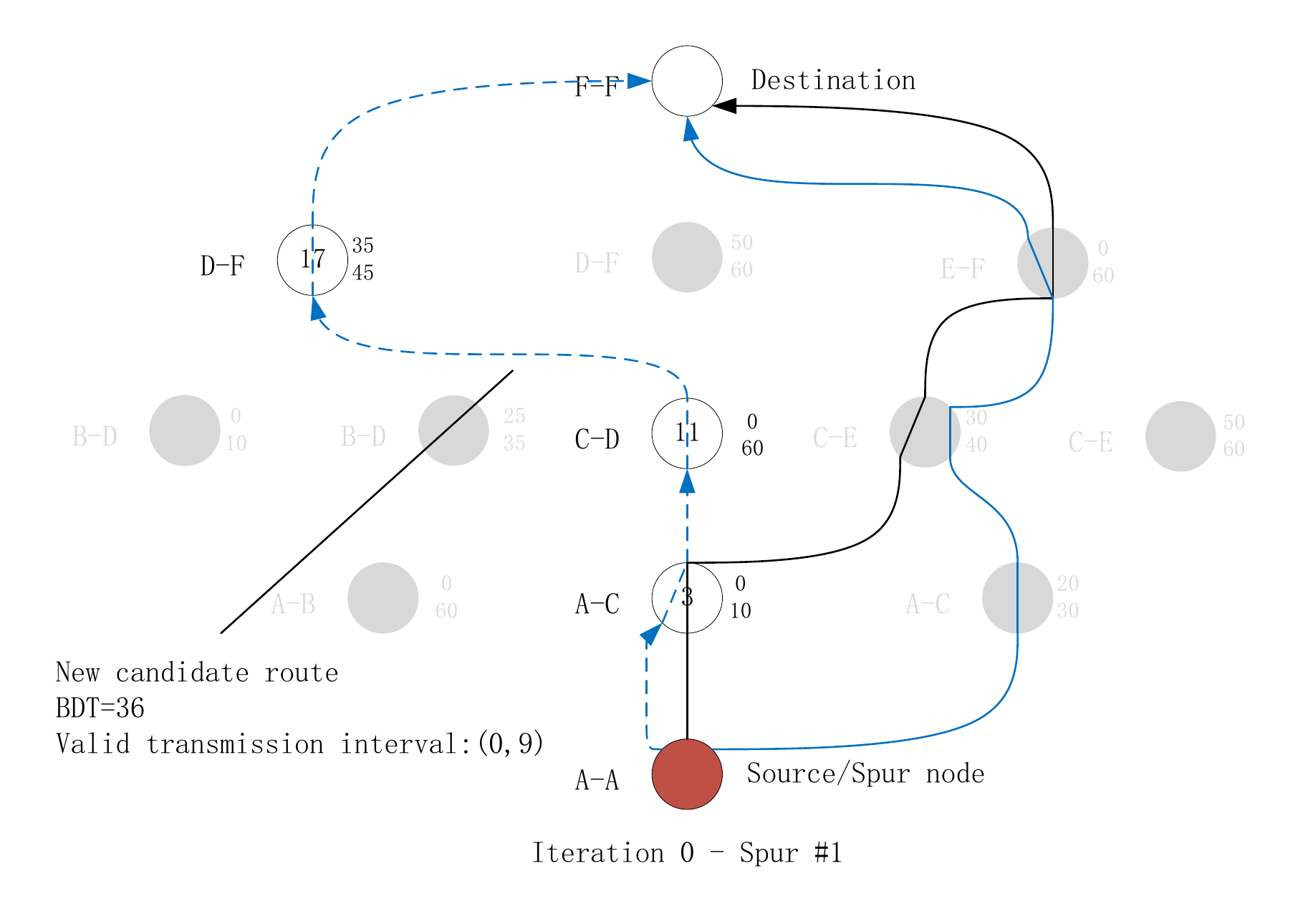}%
    }
        \\
        \subfloat[]{\includegraphics[width=0.31\textwidth]{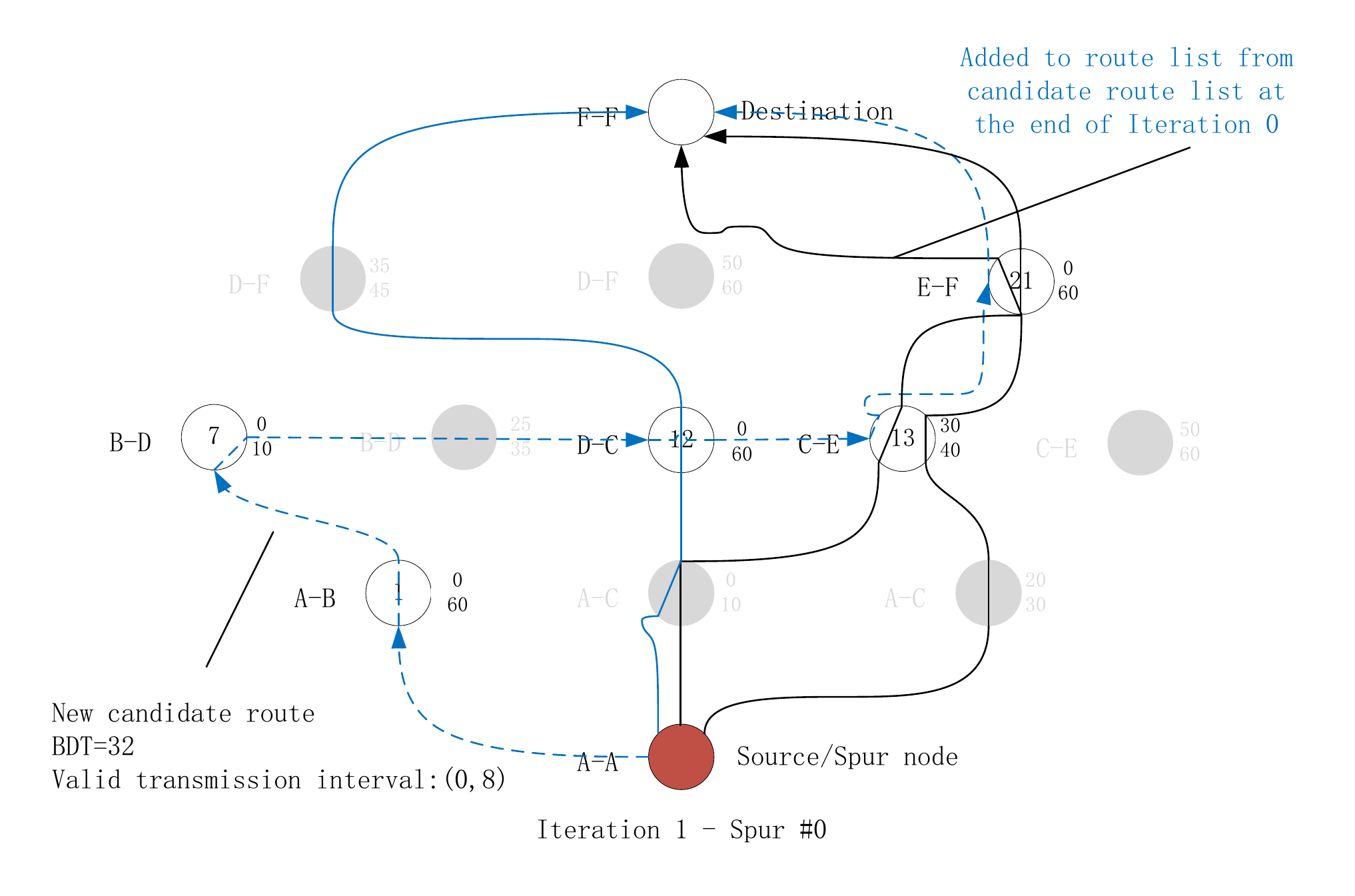}%
    }
        ~
        \subfloat[]{\includegraphics[width=0.31\textwidth]{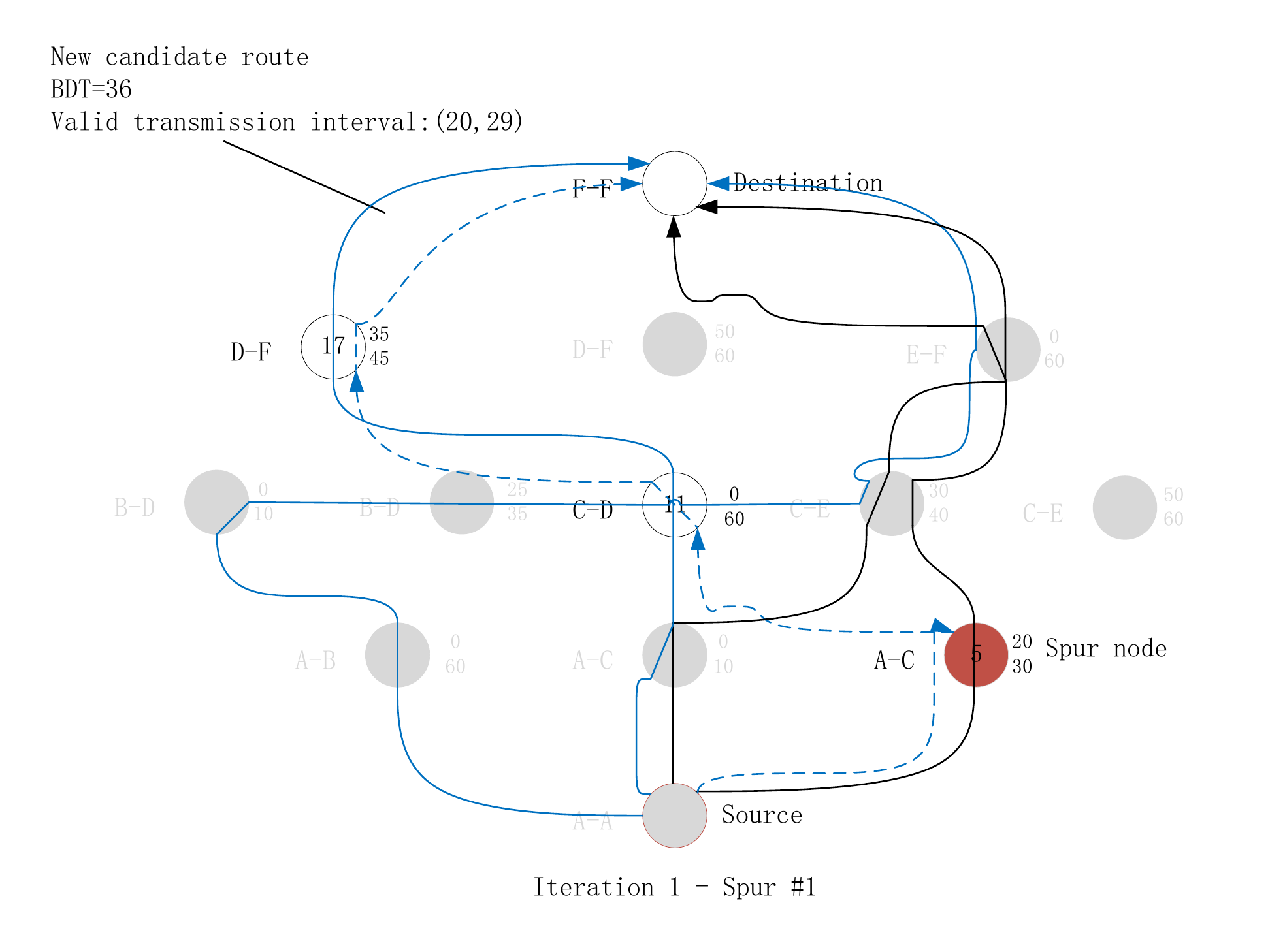}%
    }
        ~
        \subfloat[]{\includegraphics[width=0.31\textwidth]{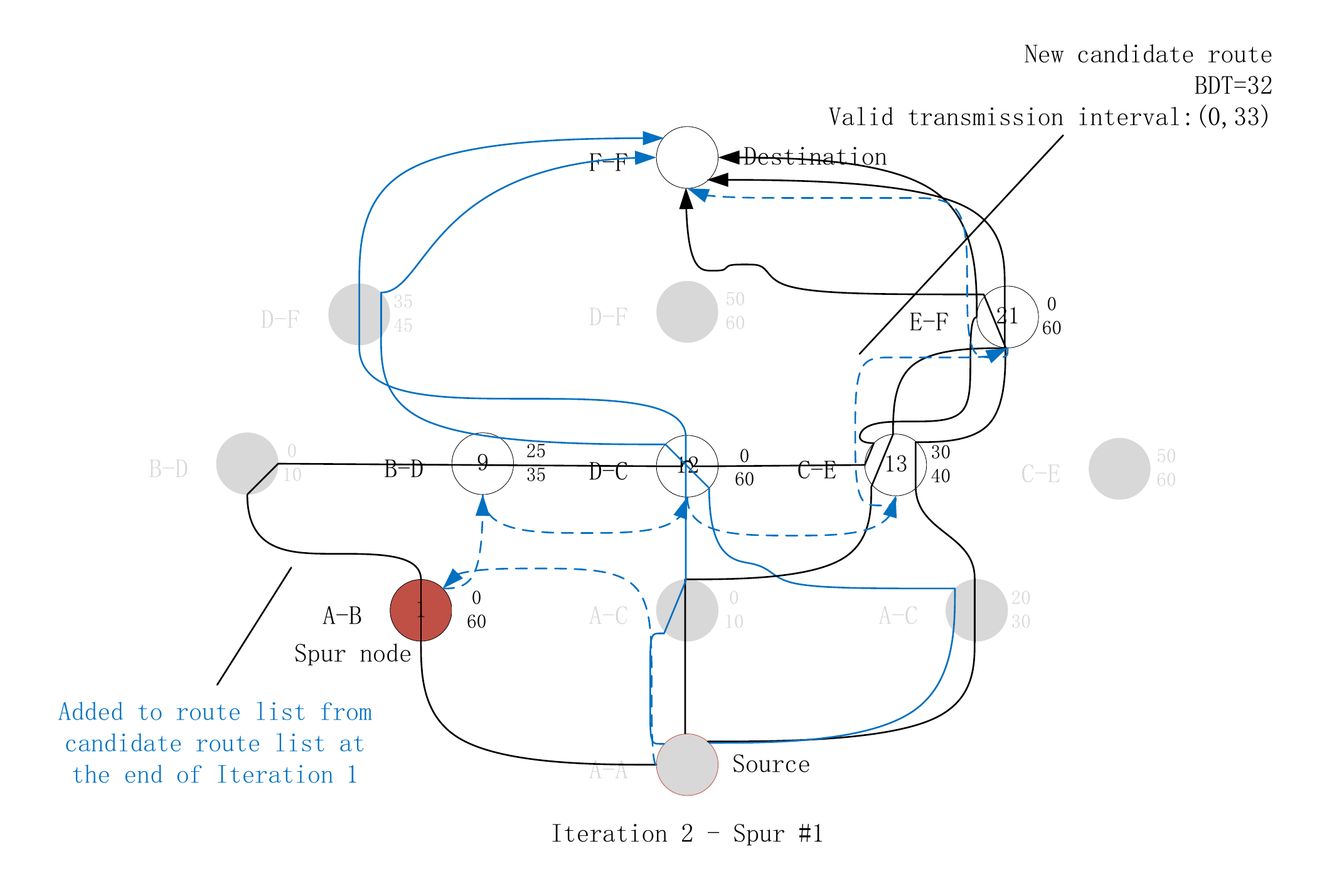}%
    }
        \\
        \subfloat[]{\includegraphics[width=0.31\textwidth]{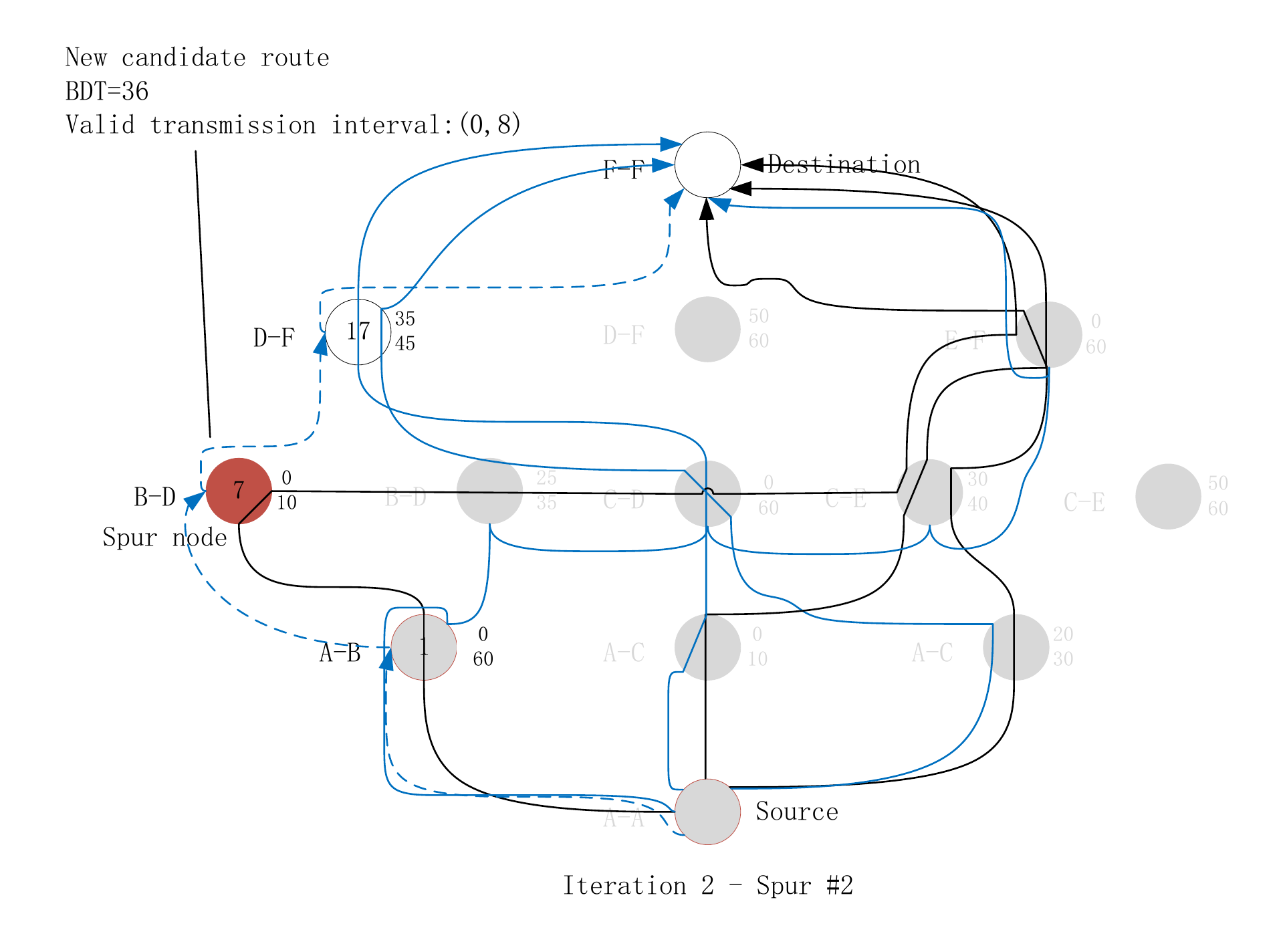}%
    }
        ~
        \subfloat[]{\includegraphics[width=0.31\textwidth]{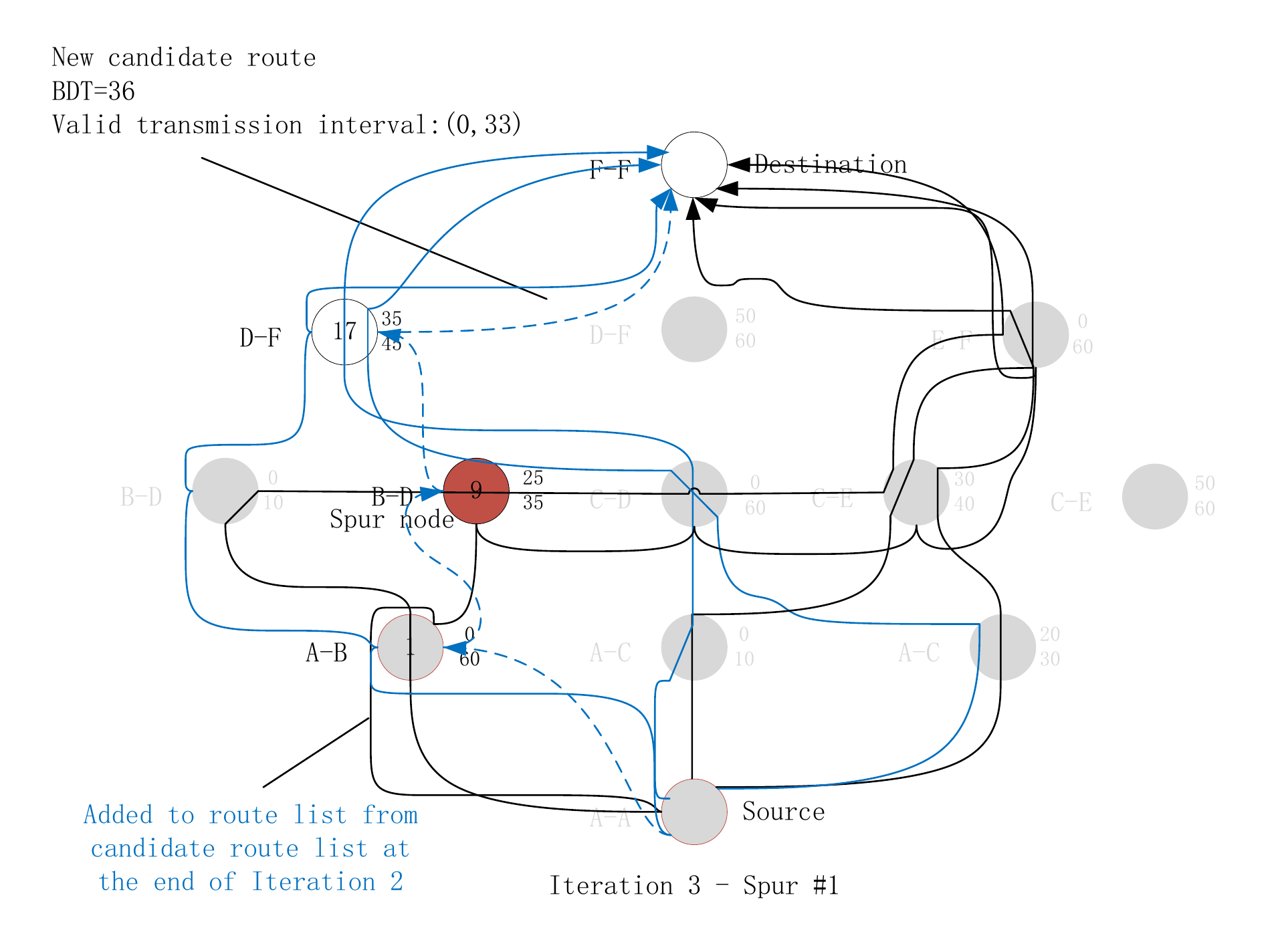}%
    }
        ~
        \subfloat[]{\includegraphics[width=0.31\textwidth]{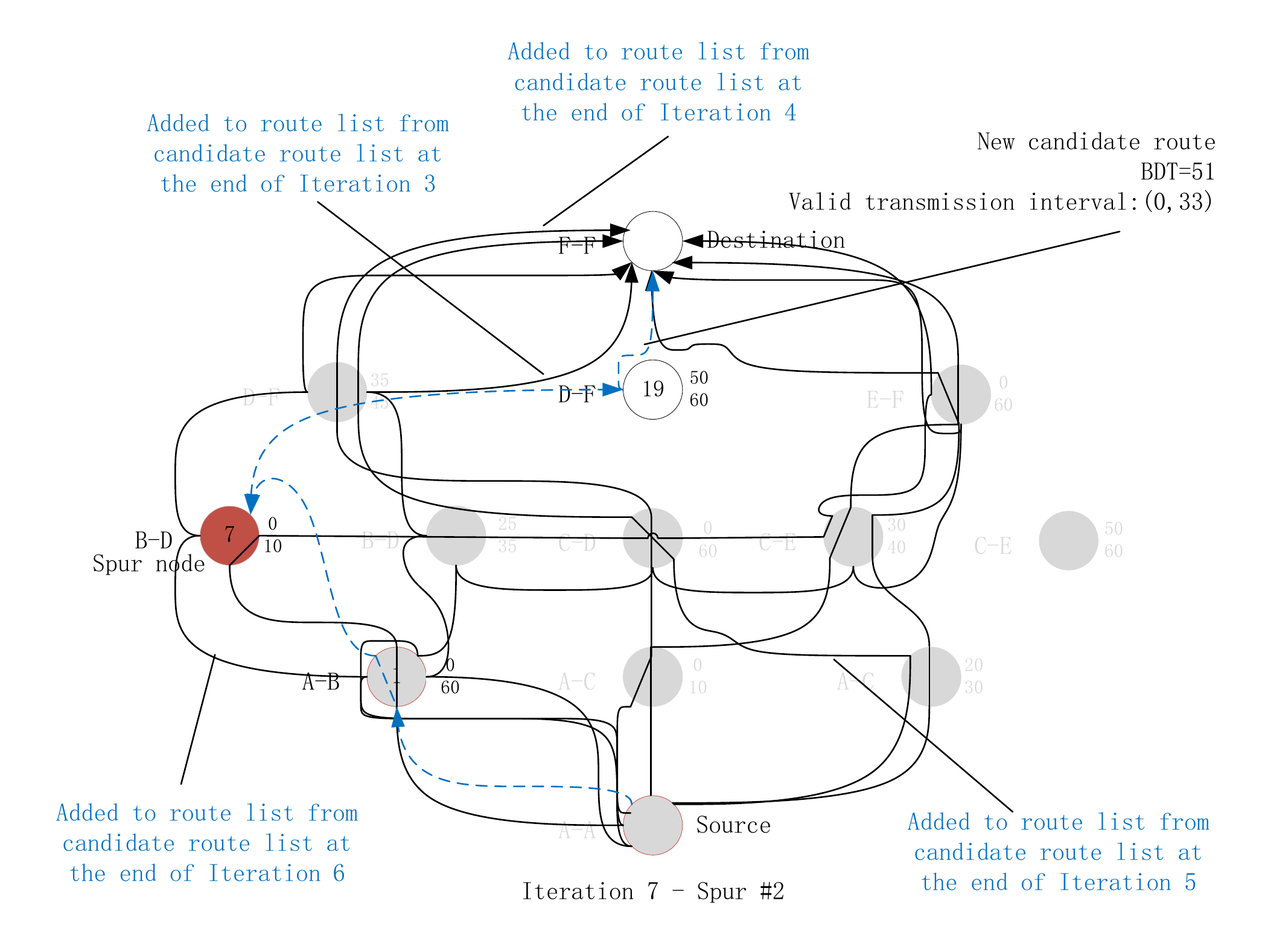}%
    }
    \caption{The iterations of the Yen-PLUS algorithm when applied to the example network in Fig.1 with K=7.}\label{Fig5}
\end{figure*}

The step-by-step iterative process and calculation results of the example network in Fig. \ref{Fig1} executing the above Yen-PLUS algorithm are shown in Fig. \ref{Fig5}. As mentioned above, the Dijkstra algorithm is executed on the whole graph for the first time in the initialization phase, and the best route $R_A^F=\{C_{A,C}^{0,10},C_{C,E}^{30,40},C_{E,F}^{0,60}\}$ is found. Then, in each iteration, the initial shortest path is the spur path, and spur nodes are replaced in turn to find new potential routes. For example, in spur$\#$0 in iteration 0,$R_A^F=\{C_{A,C}^{20,30},C_{C,E}^{30,40},C_{E,F}^{0,60}\}$ is found. Although another potential path is found in the second spur node spur$\#$1, the path found in spur$\#$0 is added to the final routing list under the cost ranking of $\mathbb{R}_i.BDT$. In this example network, the algorithm terminates after iteration 7, because a new path with $\mathbb{R}_i.BDT$ of 51 has been found, so it can be confirmed that the first K=7 is the optimal path.

Table \ref{Table1} sorts out the 9 best paths found by the above Yen-PLUS algorithm, and the detailed routes correspond to each in Fig. \ref{Fig5} and the iterations in which they were searched. It can be clearly seen that in order to determine the first 7 optimal paths, we have undergone one initialization and 8 iterations, and finally found 9 paths. It can be seen from this table that most of the shortest paths (sequence numbers 1-8) have been found when looping to iteration 3, and it can also be seen that these shortest paths are not added to the final routing list in order, which is reflected in the iteration column, it is out of order. This shows that the traditional Yen algorithm can only guarantee that the best route found in a single iteration is the smallest $\mathbb{R}_i.BDT$. Therefore, the Yen-PLUS algorithm we propose re-establishes the cost concept $\mathbb{R}_i.EDT$ in the shortest path, so that the first K of the final routing list $[R_S^D]$ must be optimal.

\begin{table*}[]
   \caption{The resulting routing list $[R_S^D]$ and its corresponding relationship sorted by $\mathbb{R}_i.EDT$ cost}
   \label{Table1}
   \centering
   \resizebox{0.85\linewidth}{!}{
   \begin{tabular}{|c|c|c|c|c|c|c|}
   \hline
   Sequence number & $\mathbb{R}_i.BDT$ & Volume & VTI     & Hops in R       & Iteration           & Corresponding graph \\ \hline
   1               & 32    & 10     & [0,9]   & 23,3,13,21      & Initialization      & (a)                  \\ \hline
   2               & 32    & 10     & [20,29] & 23,5,13,21      & Iteration 0-Spur \#0 & (b)                  \\ \hline
   3               & 32    & 10     & [0.33]  & 23,1,7,12,13,21 & Iteration 1-Spur \#0 & (d)                  \\ \hline
   4               & 32    & 9      & [0,8]   & 23,1,9,12,13,21 & Iteration 2-Spur \#1 & (f)                  \\ \hline
   5               & 36    & 10     & [0.33]  & 23,1,9,17       & Iteration 3-Spur \#1 & (h)                  \\ \hline
   6               & 36    & 10     & [0,9]   & 23,3,11,17      & Iteration 0-Spur \#1 & (c)                  \\ \hline
   7               & 36    & 10     & [20,29] & 23,5,11,17      & Iteration 1-Spur \#1 & (e)                  \\ \hline
   8               & 36    & 9      & [0,8]   & 23,1,7,17       & Iteration 2-Spur \#2 & (g)                  \\ \hline
   9               & 51    & 9      & [0,8]   & 23,1,7,19       & Iteration 7-Spur \#2 & (i)                  \\ \hline
   \end{tabular}}
   \end{table*}

Considering that this paper focuses on the large-scale dynamic network of LEO whose CP often changes, if the candidate route cannot be found for a certain bundle after the execution of the above algorithm, it will be placed in the storage of the local node. After the CP changes, it is expected the Yen-PLUS algorithm is executed again to update $[R_S^D]$ in the future until the expiration time of the bundle comes, and the bundle is deleted. If the orbital status of all satellites can be predicted during the entire cycle by using STK software, regardless of future CP changes, if the candidate route of this bundle is not found, it will never be able to reach the destination, that is, the delivery will fail.

\subsection{Dynamic Route Computation}
In this process, we adhere to the rule constructions constructed by the candidate route construction algorithm in the standard CGR\cite{r12} and adopt the method proposed in the latest implementation of ION to deal with "overbooked" a priori. When calculating ETO, a redistribution mechanism after overbooked is added to the priority in the bundle attribute\cite{r30} in order to avoid congestion and improve the successful delivery rate. In addition, if the bundle with lower priority cannot be transmitted according to the predetermined route, it will be rolled back to the upstream node for redistribution.

In addition to the above functions that have been implemented in the latest version of the ION and CGR standards, we have also innovatively added three optimization details to the RMDG-CGR algorithm:

\begin{enumerate}
\item{During the forwarding process, the original distribution of paths based on the arrival order of the bundles has changed to according to the expiration time and priority of the bundles to affect the cost function and determine the order in which the paths are allocated.}
\item{For the standard CGR mechanism, critical bundles will be copied to all candidate routes for transmission. If this method is applied to the LEO large-scale predictable constellation scenario, it will waste contact volume and affect the delivery of non-critical bundles. Therefore, we improved the critical bundle to select only the optimal paths of the downstream nodes for transmission. It will reduce the amount of invalid replication and determine whether the downstream node has received the critical bundle, and if there is one, it will not be sent to it.}
\item{Through the 6-node example shown in Fig. \ref{Fig1}, we find that for the K optimal paths that have been calculated and selected into the route list, their effective transmission capacity is not limited by the storage capacity of the node. Instead, it is controlled by the node with the shortest VTI in the TSRCG. Therefore, it is also added to the cost function of the path search as one of the influencing factors.}

\end{enumerate}

\subsection{Analysis of Algorithm Complexity}
The time complexity index is significance for routing algorithms running in a space environment, considering the limited onboard resources that can be carried on satellites, especially in future large-scale LEO scenarios. Assuming that the number of vertices in TSRCG is $V$, task $M_i$ has a total of $N$ priorities, the number of hops of the path is $hop\_cnt$, and each step is estimated according to the worst case, we can get that the time complexity of RMDG-CGR will not exceed:

\begin{equation}
   \begin{split}
   O(&2hop\_cnt^N(\frac{|M_i|}{N})^N\times T_{CRC}+\\
   &min(2hop\_cnt^N(\frac{|M_i|}{N})^N,V^2)\times T_{Yen\_PLUS})
   \end{split}
\end{equation}
Among them, $T_{Yen\_PLUS}$ represents the algorithm time complexity of Yen-PLUS, and $T_{CRC}$ represents the algorithm time complexity of candidate routes construction.

The basis of Yen's algorithm is the modified Dijkstra search for the best route through a contact graph, whose complexity is $O(|V|log(|V|))$\cite{r12}. In Yen-PLUS, many contacts are ignored because they do not meet the constraints and do not participate in iterative calls. So suppose we call modified Dijkstra search in the algorithm as $K\times l$, where $K$ is the number of routes to be found,$l$ is the average size of the spur path is $log|V|$, and we estimate that the worst-case is $|V|$, then the overall worst-case time complexity is $O(|V|log(|V|)\times K|V|)$, so it is approximately equal to $O(K|V|^2log(|V|))$. Using the worst-case estimate of $hop\_cnt$ and $T_{CRC}$, we can get that the final time complexity does not exceed:

\begin{equation}
   \begin{split}
   O(&V^N(\frac{|M_i|}{N})^N\times K|V|+\\
   &min(V^N(\frac{|M_i|}{N})^N,V^2)\times K|V|^2log(|V|))
   \end{split}
\end{equation}

\section{Performance evaluation}
In this section, we will prove the performance of our proposed RMDG-CGR strategy through comprehensive numerical results and analysis and evaluation.
\subsection{Simulation Settings}
\begin{figure*}
    \centering
            \subfloat[]{\includegraphics[width=0.35\textwidth]{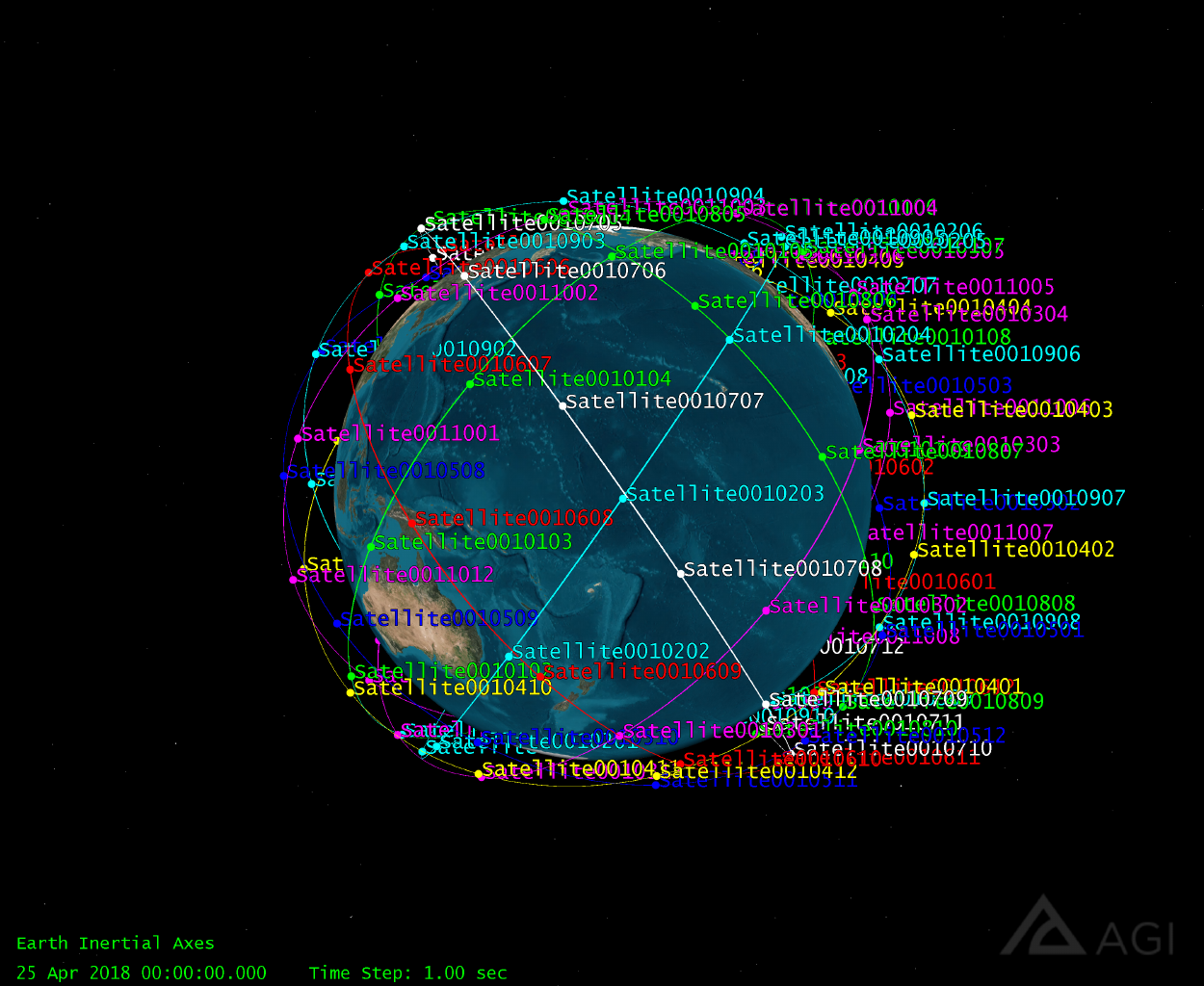}%
    }
    ~
            \subfloat[]{\includegraphics[width=0.54\textwidth]{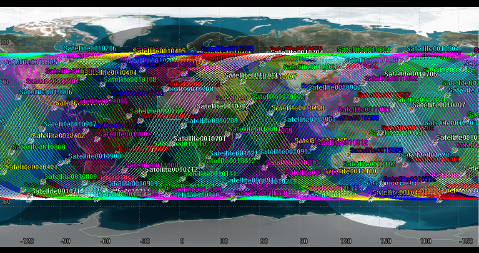}%
    }
    \caption{(a) STK simulation diagram and (b) schematic diagram of the sub-satellite point of NeLS constellation.}
    \label{Fig6}
\end{figure*}

We take the typical LEO constellation NeLS as an example to simulate a satellite network based on the DTN protocol in scenarios with multi-tasks with different priorities. Among them, the NeLS constellation is the first global satellite network planned to adopt WDM ISL technology. Its configuration is a Delta-type constellation. Its three-dimensional simulation diagram and sub-satellite point trajectory are shown in Fig. \ref{Fig6}. Using Walker constellation description method\cite{r31} can be identified as $12*10/10/1:1200:55$\cite{r20}, that is, the number of satellites in each orbital plane is 12, a total of 10 orbital planes, the phase factor is 1, the orbit height is 1200km, and the orbit inclination is 55\textdegree. The remaining specific parameters of the NeLS constellation are shown in TABLE \ref{Table2}.

We simulate the NeLS constellation in the STK, and obtain the latitude, longitude, altitude, and rate of change of each satellite per second. Then we use MATLAB to numerically analyze and calculate the CSV file output by STK, and obtain the link distance between each satellite and all other satellites per second in the NELS constellation in a constellation cycle, and the number of linkable satellites within the effective range including intra-orbits and inter-orbits, and the duration of the communicable satellite, etc. in a constellation period. Through the above MATLAB programming process, the MAT format data with the ability to generate contact graphs is obtained as the input of Python. Finally, all the algorithms proposed in this paper are written in Python and verified by the following experiments.

\begin{table}[t]
\caption{Primary Parameters of the NeLS Constellation}
\label{Table2}
\centering
\begin{tabular}{|c|c|}
\hline
Parameters                            & NeLS    \\ \hline
Orbit Inclination                     & 55\textdegree     \\ \hline
Orbital Altitude                      & 1200km  \\ \hline
Orbit Period                          & 6565s   \\ \hline
Number of Orbits                      & 10      \\ \hline
Number of Satellites Per Orbit        & 12      \\ \hline
Number of ISL Terminals Per Satellite & 4       \\ \hline
Intraorbit ISL Distance               & 3922km  \\ \hline
Interorbit ISL Distance               & $\leq$4909km \\ \hline
Inter Plane Spacing                   & 1\textdegree      \\ \hline
\end{tabular}
\end{table}

In addition, this article has the following detailed settings during the simulation process. After verification and calculation, the total transit time of any two satellites in the NeLS constellation meeting the communication distance limitation is less than 1 light-seconds according to the calculation method of OWLT in Section \uppercase\expandafter{\romannumeral2}. Therefore, according to the most pessimistic situation, we estimate the propagation delay on the way does not affect the network performance, so we uniformly set it to 1s. There are a total of 3 priority levels for bundle tasks, and related attributes are randomly generated, such as destination, data size, expiration time, and generation time. The constraints on the amount of data are described in detail in the task model part of Section \uppercase\expandafter{\romannumeral2}. It is worth noting that in all tasks, the total amount of tasks with priority 2 and 1 accounted for 25\%, and the total amount of tasks with priority 0 accounted for 75\%. The transmission rate is 1M/s, and the expiration time is set based on random numbers based on the influence of the generation time and bundle size, and the general value is 20$\sim$30s based on the generation time. In this paper, we have simulated and analyzed the situation with multiple tasks with different priorities, that is, we have set up two types of test data with and without critical bundles. Since the NeLS constellation satellites are evenly distributed as Fig.\ref{Fig6} is shown, any of them is universally selected and representative. Therefore, in both types of test data, we select the satellite node labeled 1 as the source satellite, and the destination satellite was randomly generated among 120 nodes.

\begin{figure*}[!t]
    \centering
    \includegraphics[width=\textwidth]{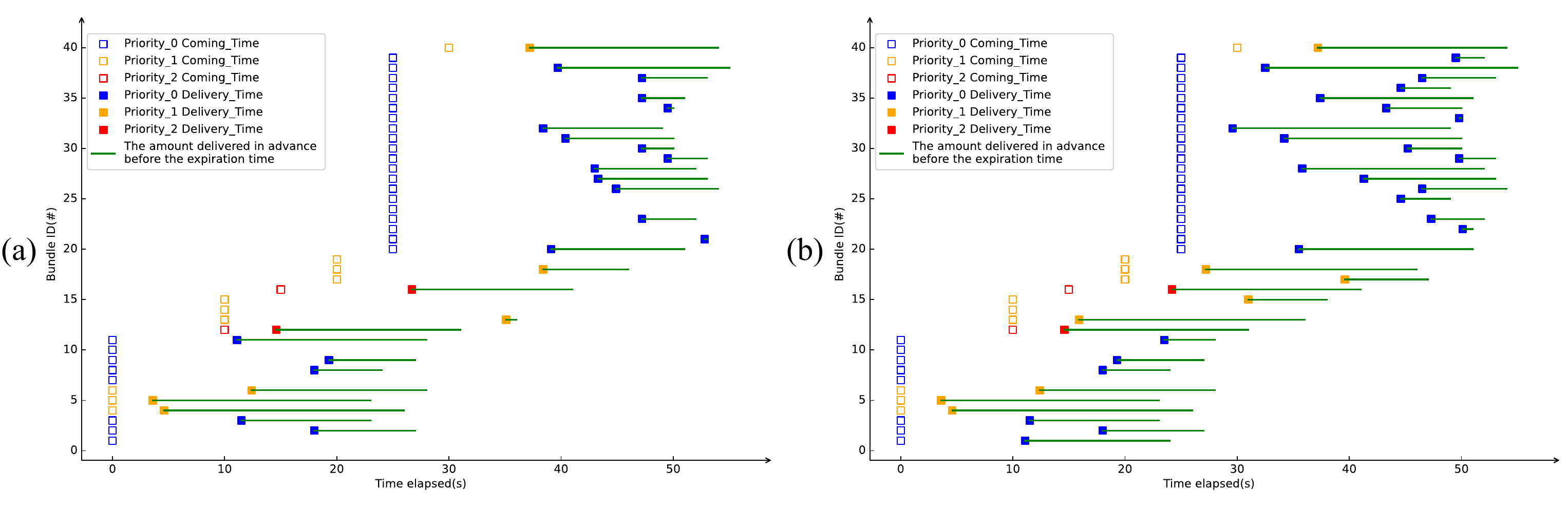}
    \caption{The delivery effect of (a) standard CGR and (b) RMDG-CGR algorithm on the transmission of bundles with streaming, expedited and data traffic.}
    \label{Fig7}
\end{figure*}

\begin{figure*}[!t]
    \centering
    \includegraphics[width=\textwidth]{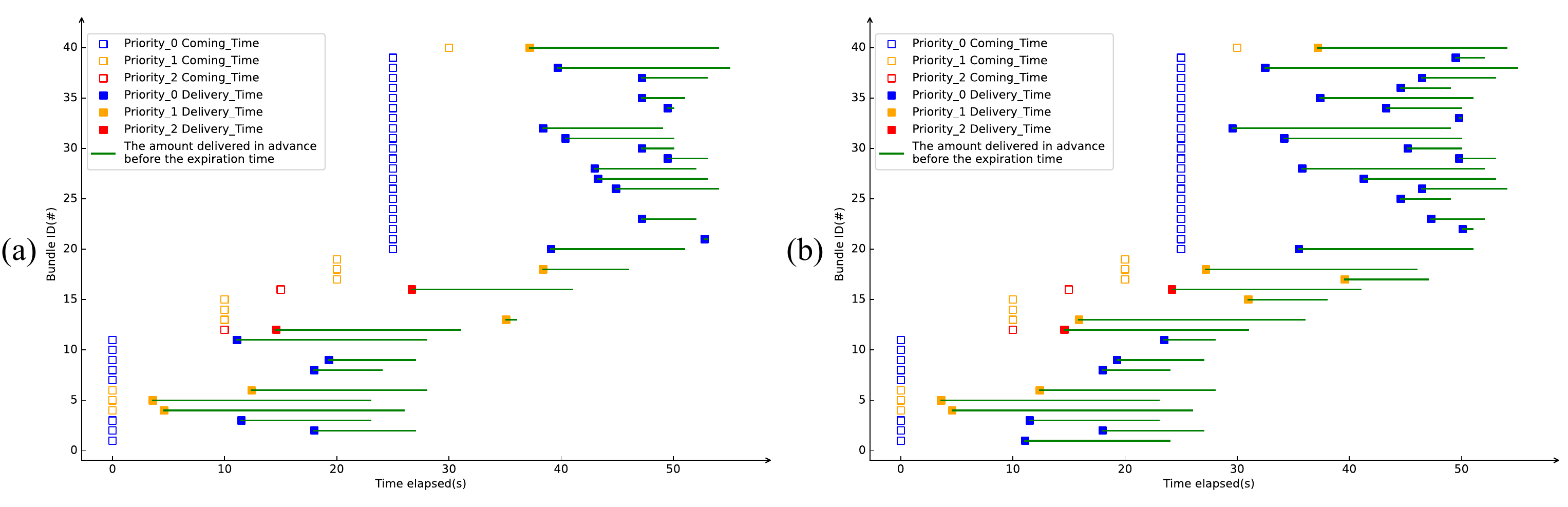}
    \caption{The delivery effect of (a)standard CGR and (b) RMDG-CGR algorithm on the transmission of bundles with expedited and data traffic.}
    \label{Fig8}
\end{figure*}

\begin{figure*}[!t]
    \centering
    \includegraphics[width=\textwidth]{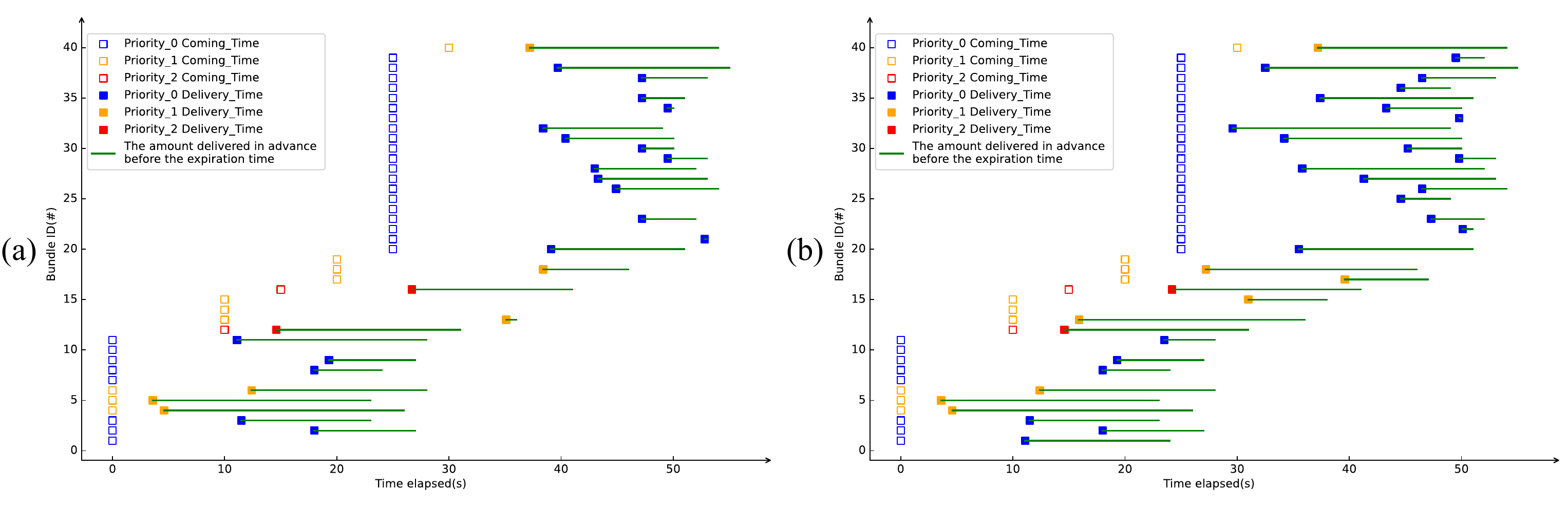}
    \caption{The contact volume occupancy rate performance of standard CGR and RMDG-CGR algorithm in two types of task scenario (a) contains streaming, expedited and data traffic bundles and (b) expedited and data traffic bundles.}
    \label{Fig9}
\end{figure*}

\subsection{Simulation Results and Discussion}
The benchmark of this article is the CGR routing search algorithm implemented in ION4.0.1. It is worth noting that this version has already covered oversubscription and the function of falling back to the upstream node to redistribute when it cannot be delivered, so the benchmark is also implemented in the experimental part of this article. In order to strictly examine the performance of our RMDG-CGR algorithm, the evaluation index basis is listed below:
(1) The amount of time for the task to be delivered in advance: the difference between the expiration time of the task and the time for successful delivery.
(2) Task successful delivery rate: the ratio of successfully delivered tasks to the total.
(3) Contact volume occupancy rate $R_O$.
(4) Satellite node routing computing resource ($S.Computing$).
(5) Satellite node storage resource ($S.Storage$).






In our simulation experiment, each indicator achieves a calculation accuracy of statistics once every 1s during the entire task transmission period. In addition, to be as accurate and rigorous as possible for the experimental results, we randomly select 20 different task models as input and perform the mean value statistics of the above indicators.

As shown in Fig. \ref{Fig7} and \ref{Fig8}, the abscissa represents the time, and the ordinate represents the bundle label. These two figures show the delivery effects of the standard CGR and RMDG-CGR algorithms for the transmission of bundles with different priorities under two task conditions. When there is a critical bundle, namely streaming traffic, it can be seen that due to the oversubscription mechanism, priority will be given to the path allocation of the critical bundle with priority 2. Therefore, the bundle of the red cube can be delivered earlier than the yellow cube with the same generation time. In Fig. \ref{Fig7} (a), it can be seen that bundle $\#$1, $\#$7, $\#$10, $\#$14, $\#$15, $\#$17, $\#$19, $\#$22, $\#$24, $\#$25, $\#$33, $\#$36 and $\#$39 are all delivery failed, and the task delivery success rate is 67.5\%. Using the same method to review the RMDG-CGR algorithm, the task delivery success rate is 85\%, which means that the task delivery rate index increased by 17.5\%. Moreover, the RMDG-CGR algorithm increases the number of tasks delivered in advance by 21/40=52.5\% compared with the standard CGR algorithm. As shown in Fig. \ref{Fig8}, when there is no critical bundle task, it can be seen that because the expedited and data types of bundles have priority 1 and 0, respectively, they do not have an oversubscription mechanism according to the settings. Therefore, it can be seen that in (a) bundle $\#$10, $\#$13, $\#$23, $\#$27, and $\#$30 all failed to be delivered. For bundle $\#$12, $\#$14, $\#$25, $\#$28, $\#$31, $\#$34 and $\#$39, the RMDG-CGR algorithm compared with the standard CGR achieves early delivery, reduces the delivery time value and increases the amount delivered in advance before the expiration time, which is the length of the green line segment shown in the figure. Therefore, it can be concluded that the RMDG-CGR algorithm we propose has better performance in terms of task delivery time and task success rate compared to standard CGR, and has a more significant improvement in scenarios with critical bundle tasks.

\begin{figure*}
    \centering
                \subfloat[]{\includegraphics[width=0.48\textwidth]{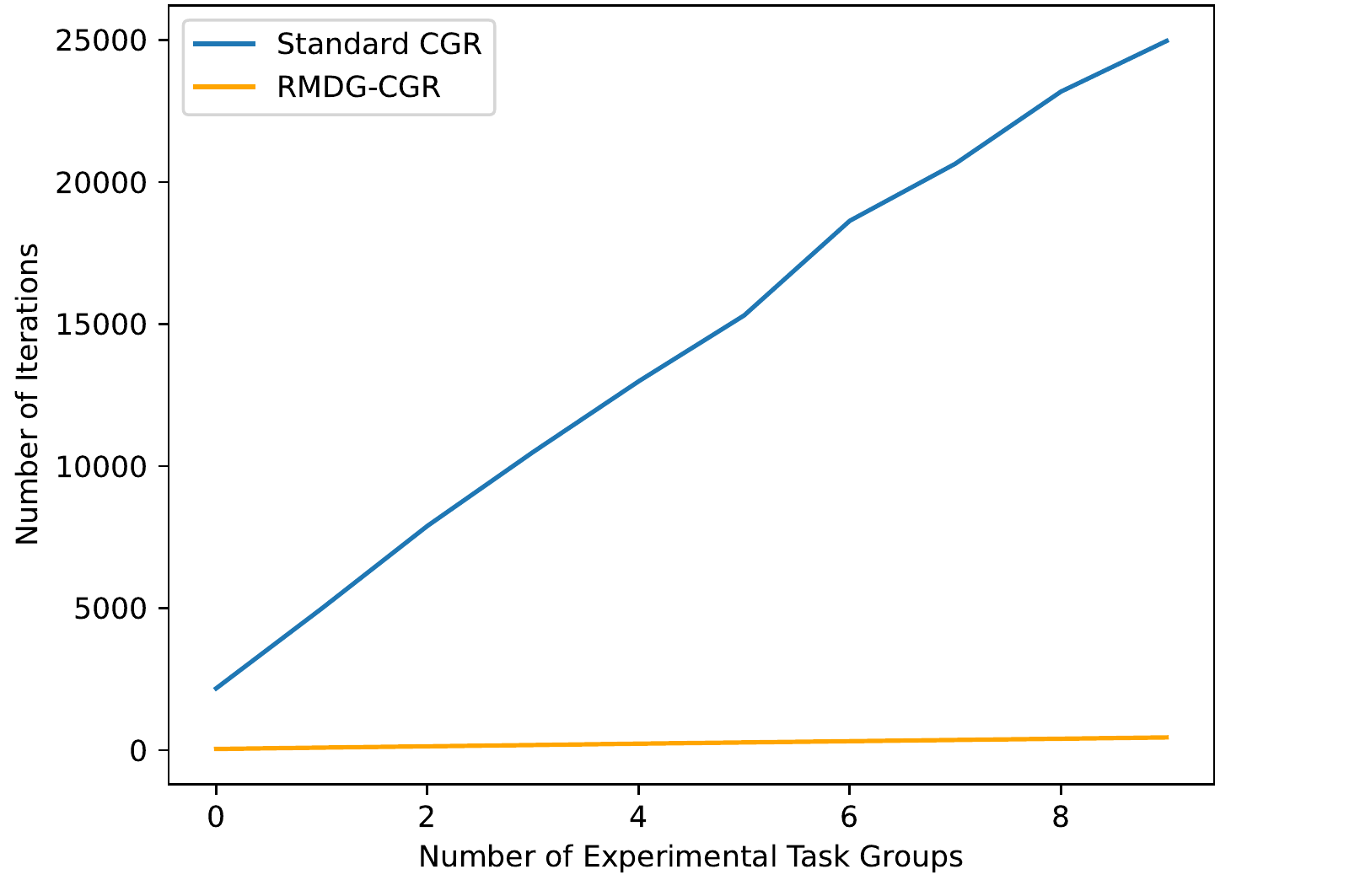}%
    }
    ~
                \subfloat[]{\includegraphics[width=0.48\textwidth]{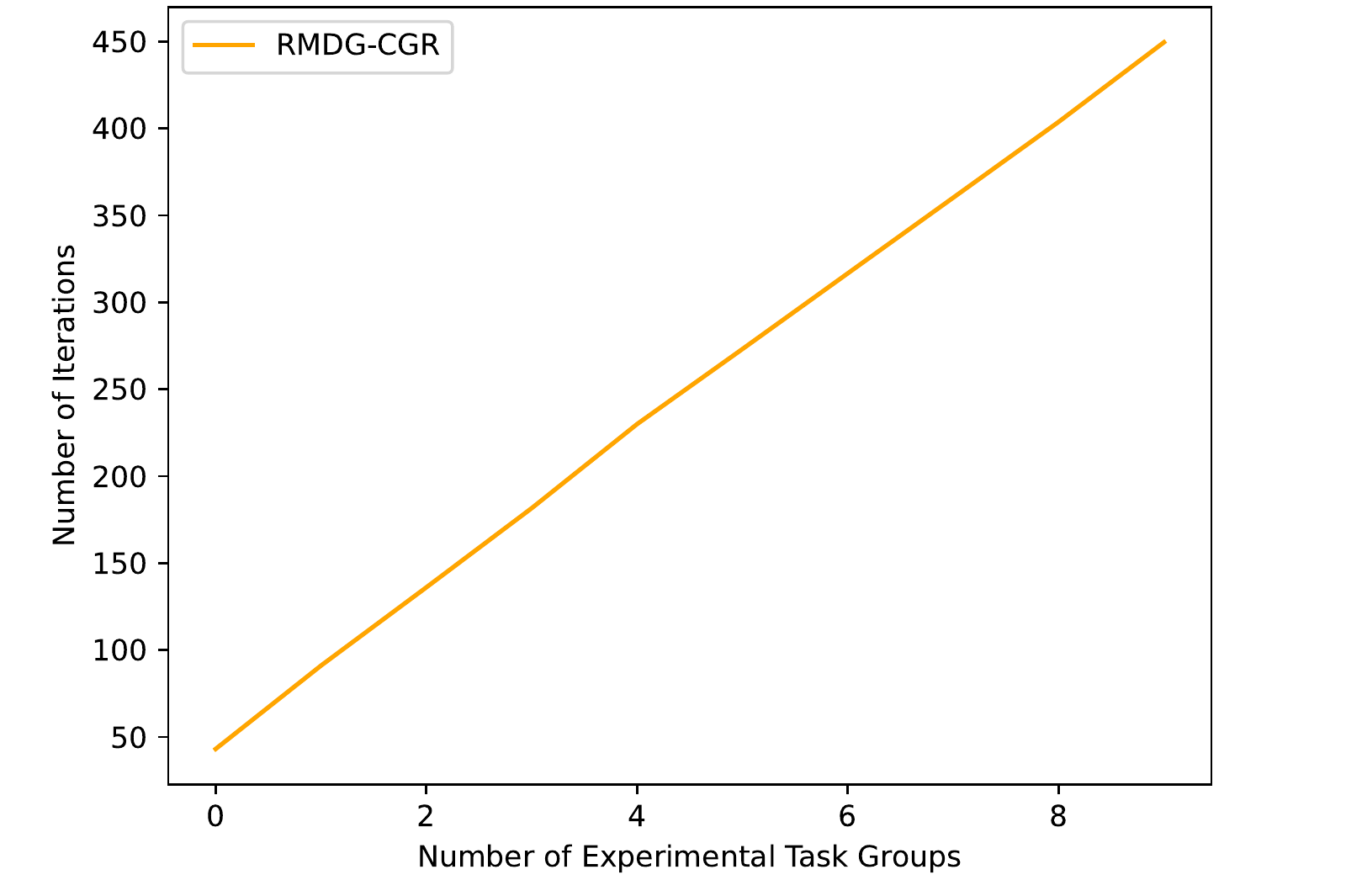}%
    }
    \caption{The computing resource consumption performance of standard CGR and RMDG-CGR algorithm with streaming, expedited and data traffic bundles.}
    \label{Fig10}
\end{figure*}

\begin{figure}
\centering
\includegraphics[width=1\linewidth]{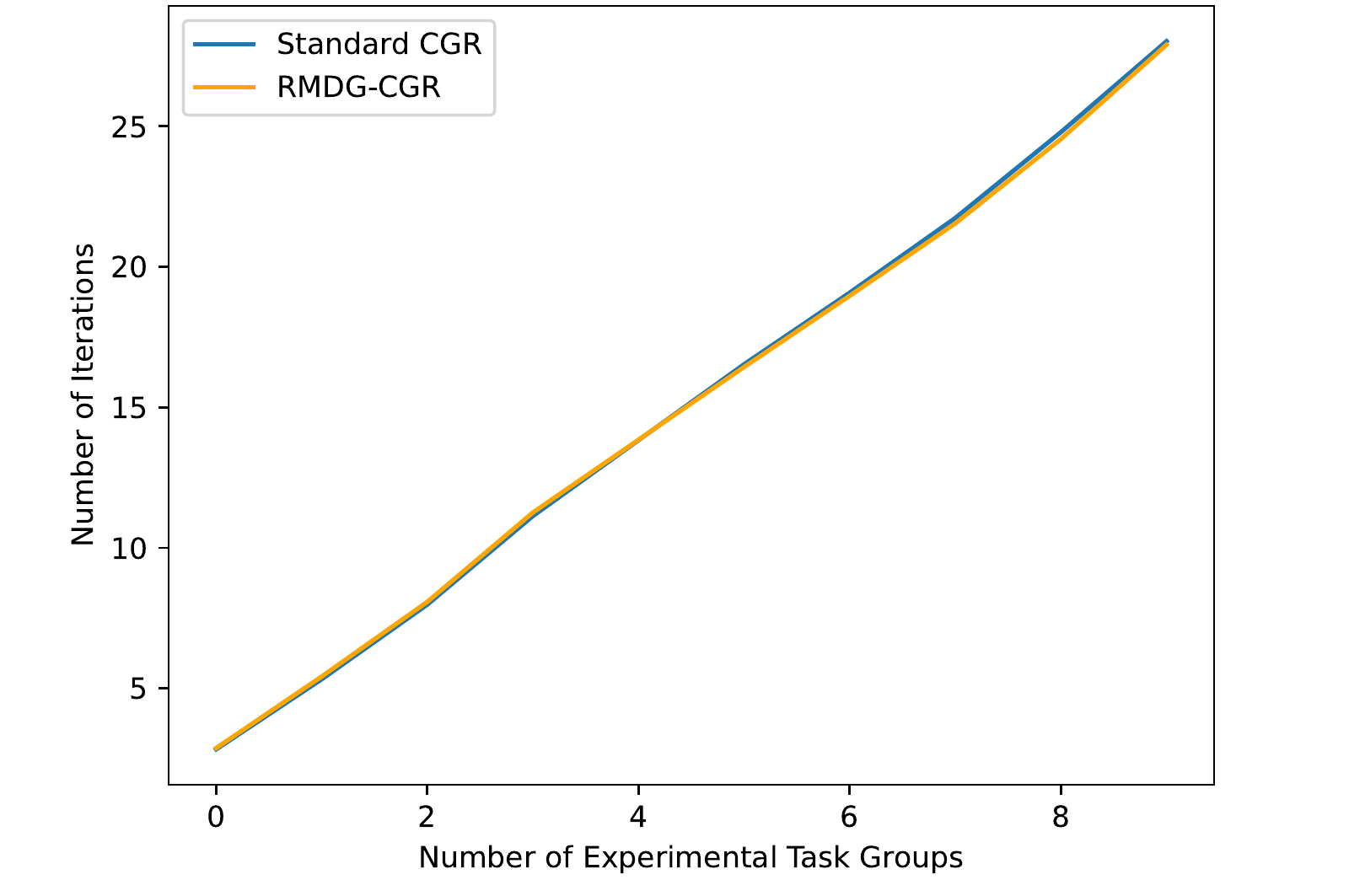}
\caption{The computing resource consumption performance of standard CGR and RMDG-CGR algorithm with expedited and data traffic bundles.}
\label{Fig11}
\end{figure}

Next, to further prove the performance of our proposed method, we compared it with the standard CGR method in terms of contact volume occupancy rate, computing resource consumption, and storage resource consumption. For evaluating the above three performance indicators, we still consider two different task plans. The first task plan contains three different priority task types, and the second task plan contains two different priority task types. The important difference is whether there is a bundle with priority 2, that is, a critical bundle.

As shown in Fig. \ref{Fig9} (a) and (b), we respectively evaluate the contact volume occupancy rate of the standard CGR and RMDG-CGR algorithms in the two task scenarios, where the ordinate data is calculated every moment. It can be seen from Fig. \ref{Fig9}(a) that in the presence of critical bundles tasks, the contact volume occupancy rate of the RMDG-CGR algorithm has been lower than the standard CGR during the entire task transmission, and the gap is about 50\%. In another task scenario, it can be seen that the performance of the two algorithms is similar, and the difference is almost no more than 0.5\%. Compared with the standard CGR algorithm, the RMDG-CGR proposed in this paper can guarantee the delivery of high-priority tasks while significantly reducing the occupancy rate of contact capacity. Aiming at the satellite network environment with limited resources, the improvement of this performance is significant.

Fig. \ref{Fig10} and \ref{Fig11} evaluate the consumption of computing resources under different task scenarios. Among them, the abscissa is the number of experiments, and the ordinate is the accumulation of the average of the number of algorithm iterations during the task transmission period corresponding to each experiment with a period of 1s. It can be seen from Fig. \ref{Fig10} that when a given task includes streaming traffic, that is, critical bundles, due to the standard CGR's strategy of copying critical bundles in all candidate routes, it can be seen that this mechanism consumes computing resources very much. In (a), the number of iterations of RMDG-CGR has been significantly lower than that of the benchmark. After 10 sets of experiments, the gap can reach about 25,000 times. This can be explained by the principle that we have improved the forwarding mechanism of critical bundles to limit their invalid replication. In order to better show the effect of RMDG-CGR separately, we have changed the ordinate scale in (b). It can be seen that the average number of iterations of a single experiment is less than 50 times, and the cumulative number of 10 experiments is about 450 times. The computational resource consumption of our proposed RMDG-CGR is only about 1.8\% (450/25000) of the standard CGR. Fig. \ref{Fig11} shows the performance of computing resource consumption under the task scheme without critical bundles. The performance of the two algorithms is similar and tends to be constant.

\begin{figure*}
    \centering
    \includegraphics[width=\textwidth]{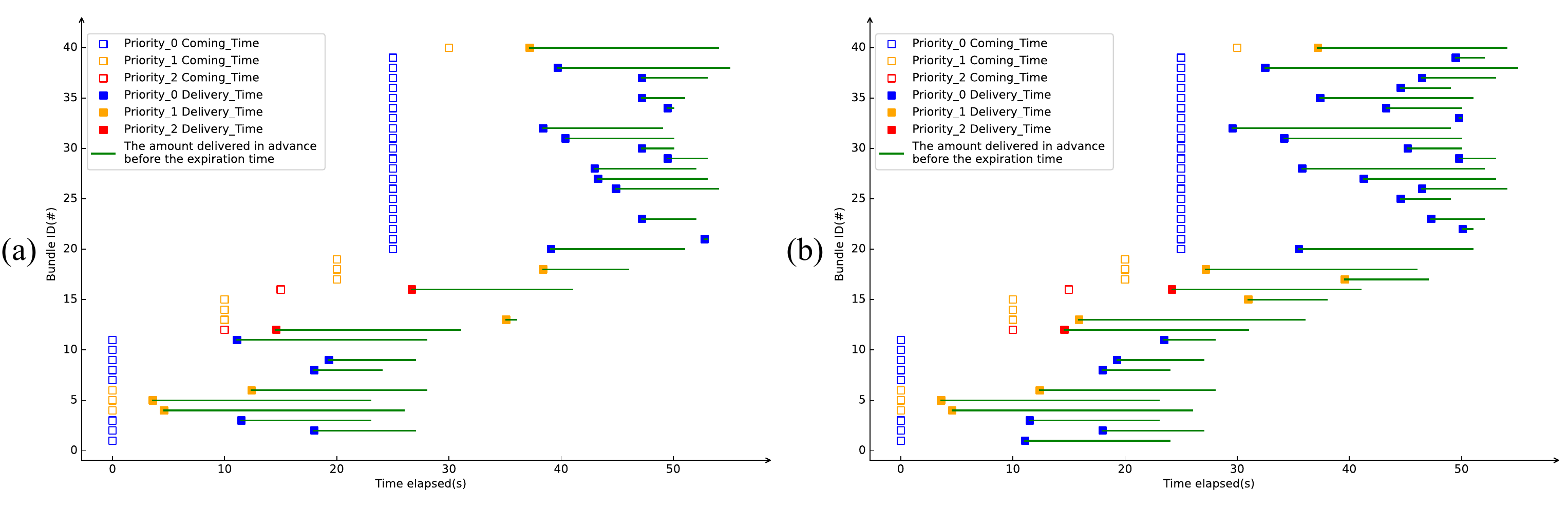}
    \caption{Changes of bundle sizes in three states on all nodes during task transmission under (a)standard CGR and (b) RMDG-CGR algorithm with streaming, expedited and data traffic.}
    \label{Fig12}
\end{figure*}

\begin{figure*}
    \centering
    \includegraphics[width=\textwidth]{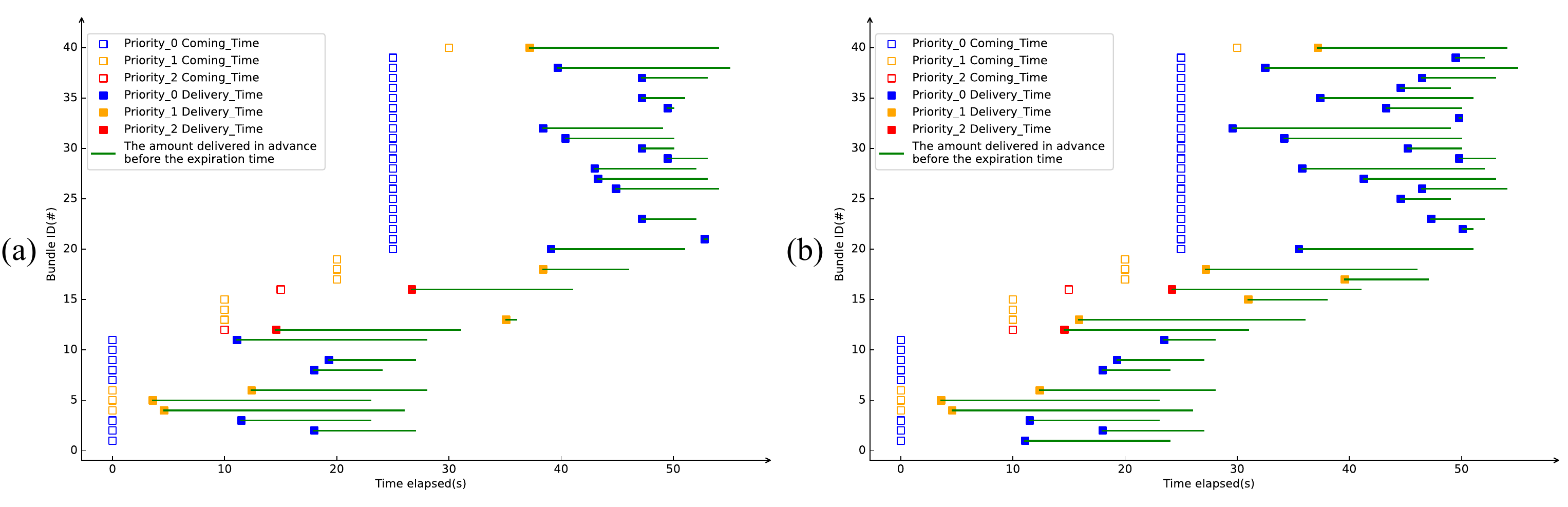}
    \caption{Changes of bundle sizes in three states on all nodes during task transmission under (a)standard CGR and (b) RMDG-CGR algorithm with expedited and data traffic.}
    \label{Fig13}
\end{figure*}

Fig. \ref{Fig12} and \ref{Fig13} respectively show the changes in the size of the bundles to send, at sending, and sent overtime during the mission transfer period of the standard CGR and RMDG-CGR algorithms under the input of the two task schemes. We can especially pay attention to the change of the orange-yellow line, which reflects the storage resource consumption of those bundles that are at sending in the satellite node, reaching the peak in the middle of the task. As the delivery of some tasks is successful, the delivery of some tasks fails, and the expiration time of some tasks expires, the final value of the orange line segment decreases to 0. It can be seen that under the input of the first task plan, the RMDG-CGR algorithm shows a better effect than the standard CGR in saving the storage capacity, reducing the storage resource consumption from more than 8,000 Mb to more than 1,200 Mb. Because the RMDG-CGR algorithm optimizes the redistribution path process of the copy of the critical bundle, greatly reducing its invalid copy. And the two algorithms perform similarly in the absence of the critical bundle.

\section{Conclusion}
In this paper, we study the resource-optimized routing strategy for multi-task delivery assurance based on large-scale LEO constellation satellite networks. Using the TSRCG, we can accurately characterize the time-varying but predictable characteristics of satellite networks from both time and space dimensions and network resource parameters under multi-tasks, such as contact volume, computing resources, storage resources, etc. Then, based on the TSRCG and multi-task model, we propose an RMDG-CGR algorithm to ensure multi-task delivery and reduce resource consumption. This method modifies the route-list computation and dynamic route computation in the bundle routing process and optimizes the route allocation mechanism for the EDT-based cost function and critical bundle in the forwarding process. In addition, theoretical analysis and experimental simulation have also been carried out to verify the algorithm. Compared with standard CGR, RMDG-CGR can achieve higher task delivery in advance and a successful task delivery rate, as well as saving contact volume occupancy rate, computing, and storage resource consumption. Especially in mission scenarios with critical bundles, the effect is more prominent.

To fairly evaluate the contribution of this article and the content that needs to be improved in the future, the advantage of our solution is the TSRCG combined multi-task model and the algorithm complexity of RMDG-CGR can be easily applied to any large-scale network with predictable topology, and when there are critical bundles in the task better algorithm performance will be obtained. However, how the Yen-PLUS algorithm in this paper controls the number of optimal routes to be calculated and how to effectively partially update the routing table without pruning when the 
CP is changed is a problem to be solved. In addition, the current routing search algorithm is still based on the Dijkstra plus greedy algorithm as the core method, and the result obtained is that the optimal at the time is not the global optimal\cite{r17}. Therefore, the above are all critical DTN routing issues that we need to study in the future under the constraints of limited network resources.

\section*{Acknowledgments}
This work is supported by the National Key Research and Development Program of China (NO.2020YFB1806000).



\bibliographystyle{IEEEtran}


\newpage

\section*{Biography Section}


\begin{IEEEbiography}[{\includegraphics[width=1in,height=1.25in,clip,keepaspectratio]{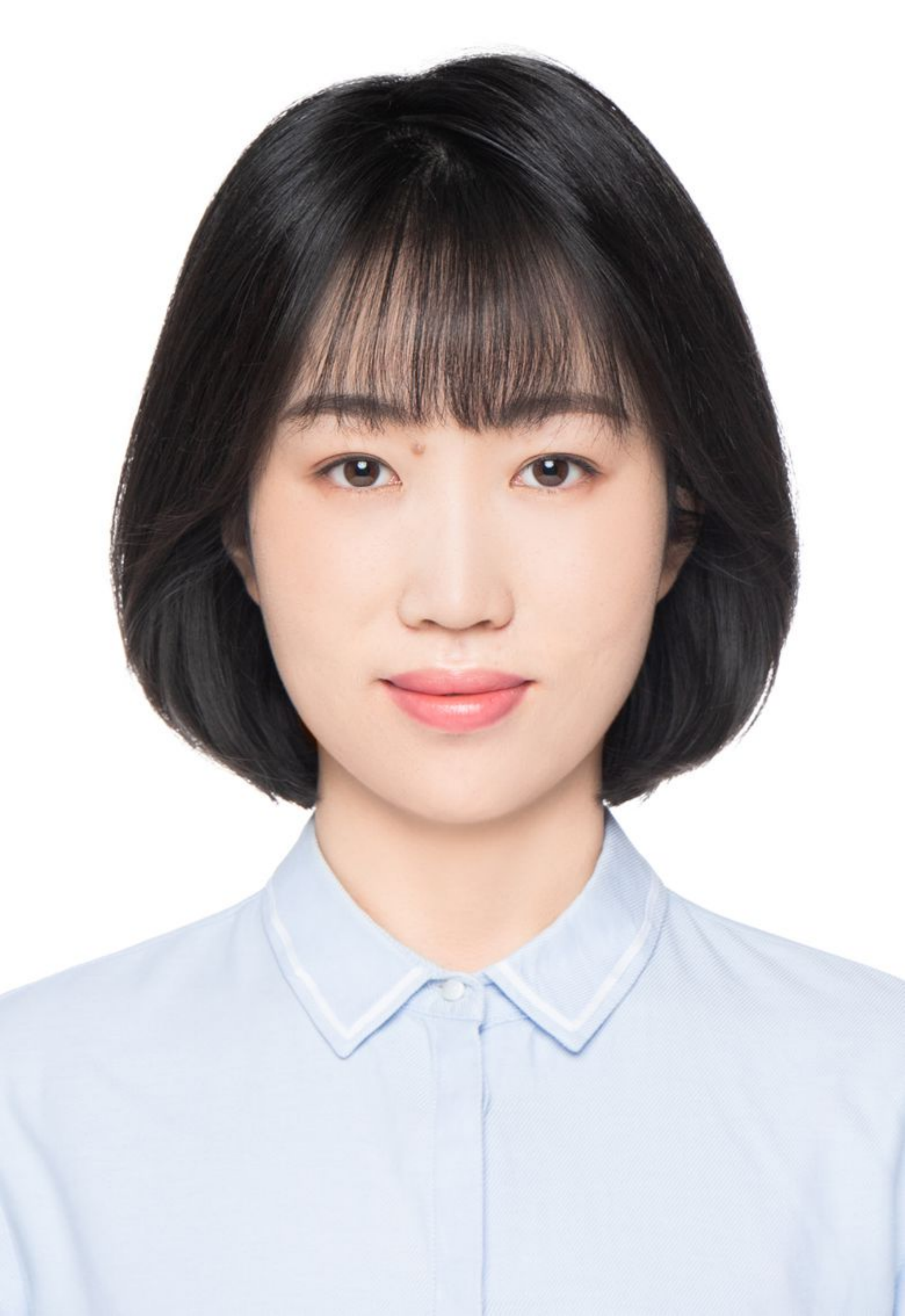}}]{Xue Sun}
received the B.S. degree in the school of Automation in 2016 from Beijing University of Posts and Telecommunications, Beijing, China. She is currently working toward the Ph.D. degree in computer application technology with the University of Chinese Academy of Sciences, Beijing, China. Her research interests include space optical communication and satellite network.
\end{IEEEbiography}


\begin{IEEEbiography}[{\includegraphics[width=1in,height=1.25in,clip,keepaspectratio]{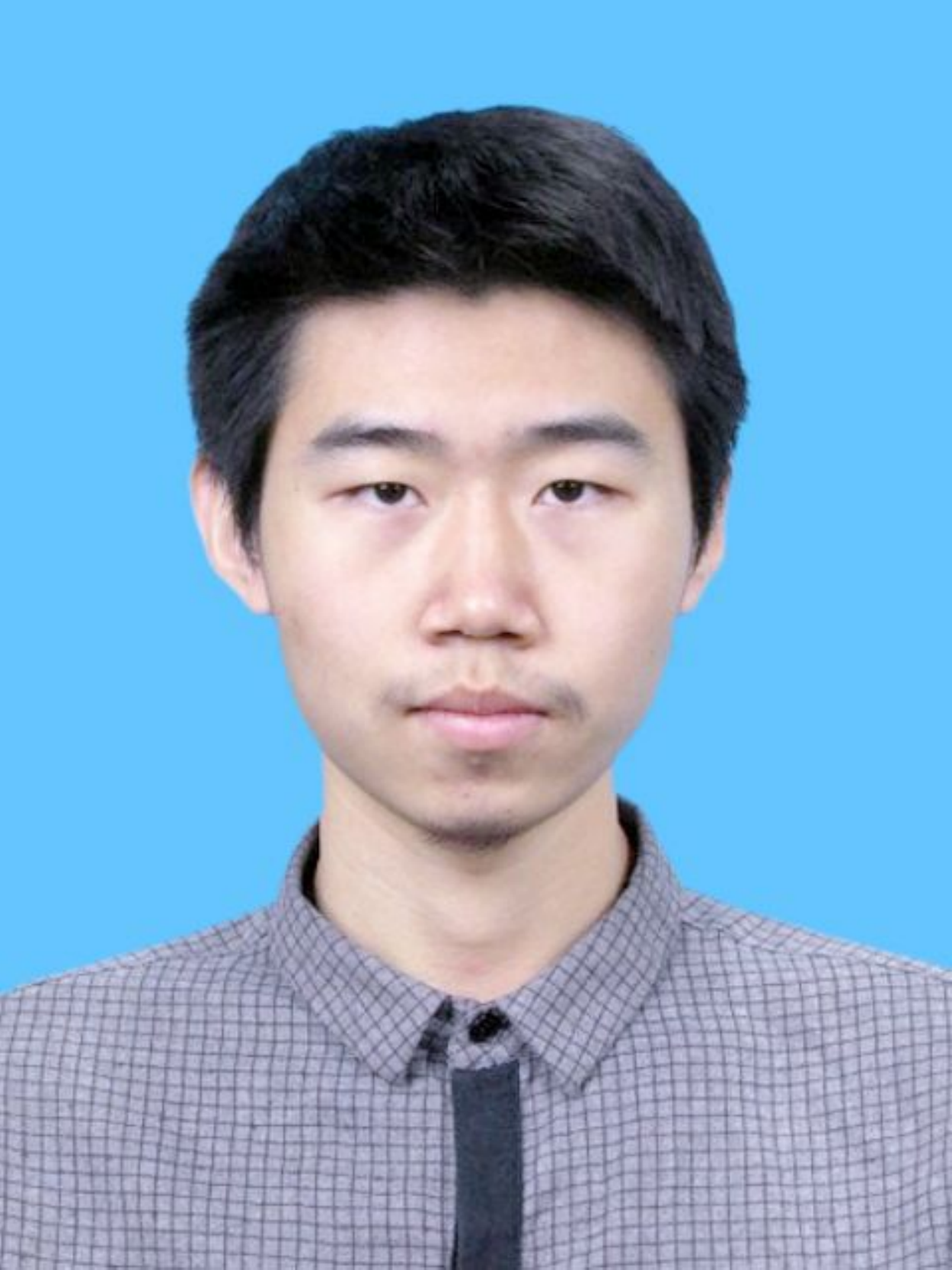}}]{Changhao Li}
received a bachelor's degree from North University of China in 2020. He is currently working towards the master’s degree in the School of Aeronautics and Astronautics, University of Chinese Academy of Sciences. His current research is routing algorithm.
\end{IEEEbiography}

\begin{IEEEbiography}[{\includegraphics[width=1in,height=1.25in,clip,keepaspectratio]{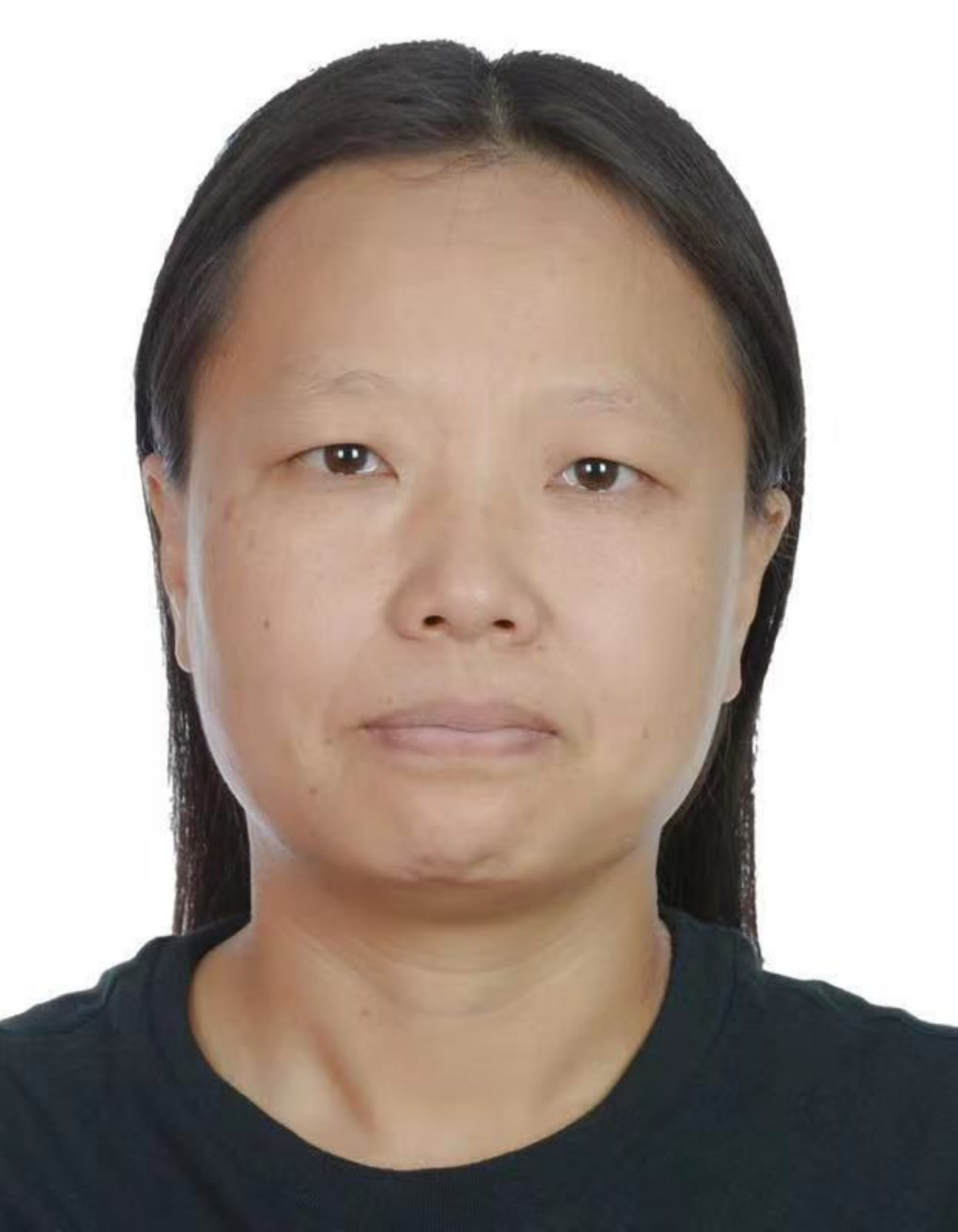}}]{Lei Yan}
received the master’s degree from Center for Space Science and Applied Research, Chinese Academy of Sciences in 2004. She’s currently a senior engineer at Technology and Engineering Center for Space Utilization, Chinese Academy of Sciences. Her research interests include satellite network, distributed computing, distributed storage and edge computing.
\end{IEEEbiography}

\begin{IEEEbiography}[{\includegraphics[width=1in,height=1.25in,clip,keepaspectratio]{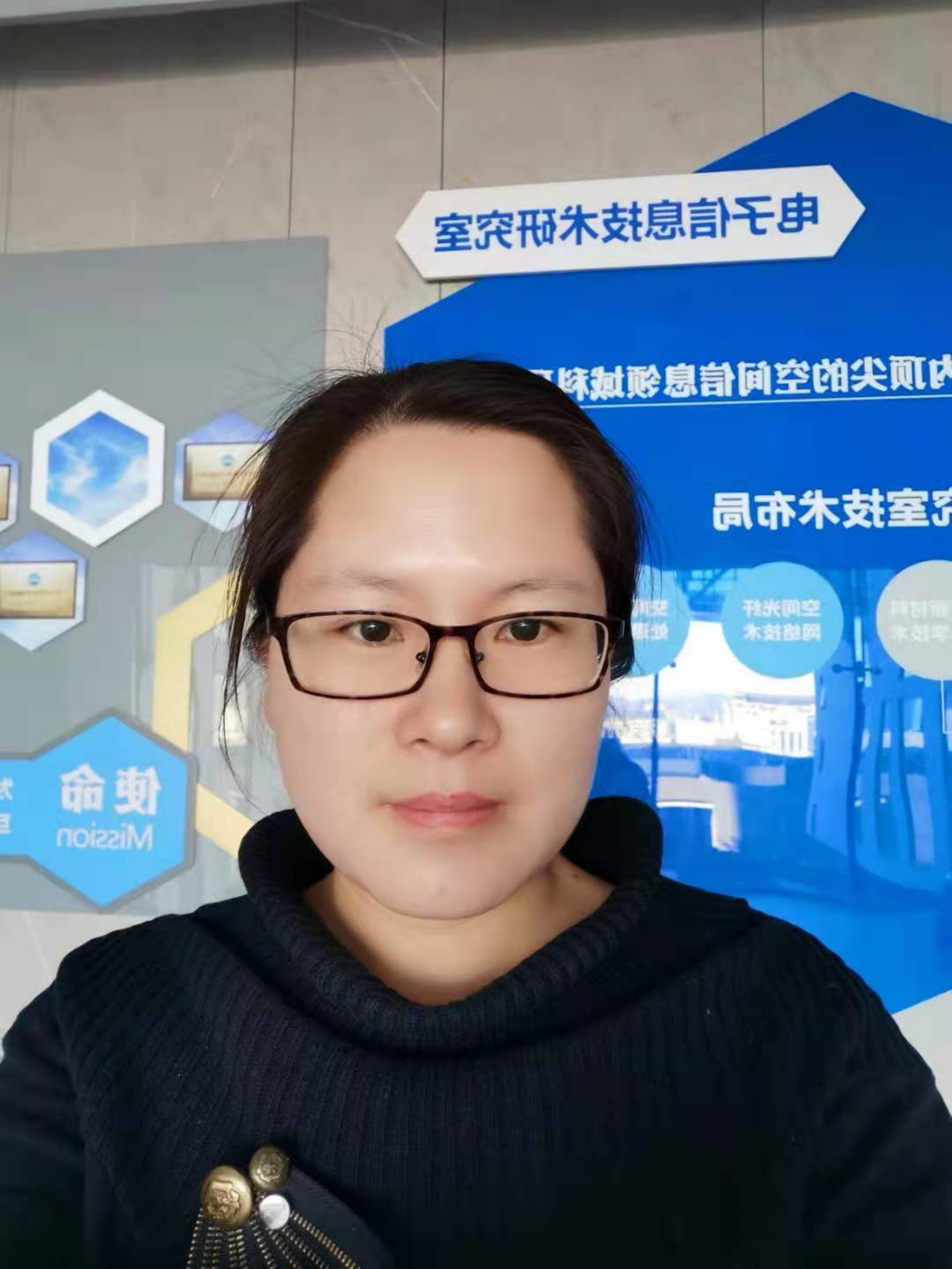}}]{Suzhi Cao}
received the bachelor’s degree and the master’s degree from Tianjin University in 2004 and 2007 respectively. She received the PhD degree from Academy of Opto-electronics, Chinese Academy of Sciences in 2010. She’s currently an associate researcher at Technology and Engineering Center for Space Utilization, Chinese Academy of Sciences. Her research interests include satellite network, edge computing and distributed computing.
\end{IEEEbiography}

\vfill

\end{document}